\pdfoutput=1
\documentclass[fleqn,usenatbib]{mnras}

\usepackage{newtxtext,newtxmath}

\usepackage[T1]{fontenc}
\usepackage{ae,aecompl}
\usepackage[normalem]{ulem}

\usepackage{graphicx}	
\usepackage{amsmath}	
\usepackage{amssymb}	
\usepackage{longtable}  

\usepackage[utf8]{inputenc}

\title[Lithium depletion in solar twins]{The Li-age correlation: the Sun is unusually Li deficient for its age
\thanks{Based on observations collected at the European Organisation
for Astronomical Research in the Southern Hemisphere under ESO
programs 188.C-0265, 183.D-0729, 292.C-5004, 097.C-0571, 092.C-
0721, 093.C-0409, 072.C-0488, 183.C-0972, 091.C-0936, 192.C-
0852, 196.C-1006, 076.C-0155, 096.C-0499, 185.D-0056, 192.C-0224,
075.C-0332, 090.C-0421, 091.C-0034, 077.C-0364, 089.C-0415, 60.A-
9036, 092.C-0832, 295.C-5035, 295.C-5031, 60.A-9700, 289.D-5015,
096.C-0210, 086.C-0284, 088.C-0323, 0100.D-0444, and 099.C-0491.}}

\author[M. Carlos et al.]{
M. Carlos$^{1}\thanks{marilia.carlos@usp.br}$, J. Mel\'endez$^{1}$, L. Spina$^{2}$, L. A. dos Santos$^{3}$, M. Bedell$^{4}$, I. Ramirez$^{5}$,\newauthor
M. Asplund$^{6}$, J. L. Bean$^{7}$, D. Yong$^{6}$, J. Yana Galarza$^{1}$,  and A. Alves-Brito$^{8}$
\\
$^{1}$Departamento de Astronomia, IAG, Universidade de S\~ao Paulo, Rua do Mat\~ao 1226 S\~ao Paulo, 05509-900, Brazil\\
$^{2}$Monash Centre for Astrophysics, School of Physics and Astronomy, Monash University, VIC 3800, Australia\\
$^{3}$Observatoire de l'Universit\'e de Gen\'eve, 51 chemin des Maillettes, CH-1290 Versoix, Switzerland\\
$^{4}$Center for Computational Astrophysics, Flatiron Institute, 162 5th Ave, New York, NY 10010, USA\\
$^{5}$Tacoma Community College, 6501 South 19th Street, Tacoma, WA 98466-7400, USA\\
$^{6}$Research School of Astronomy and Astrophysics, The Australian National University, Cotter Road, Canberra, ACT 2611, Australia\\
$^{7}$Department of Astronomy \& Astrophysics, 5640 S. Ellis Ave, Chicago, IL 60637, USA\\
$^{8}$Universidade Federal do Rio Grande do Sul, Instituto de F\'isica, Av. Bento Gon\c{c}alves 9500, Porto Alegre, RS, Brazil\\
}

\date{Accepted XXX. Received YYY; in original form ZZZ}

\pubyear{2018}

\begin{document}
\label{firstpage}
\pagerange{\pageref{firstpage}--\pageref{lastpage}}
\maketitle

\begin{abstract}
The present work aims to examine in detail the depletion of lithium in solar twins to better  constrain stellar evolution models and investigate  its possible connection with exoplanets. We employ spectral synthesis in the region of the asymmetric 6707.75 \AA \, Li I line for a sample of 77 stars plus the Sun. As in previous works based on a smaller sample of solar twins, we find a strong correlation between Li depletion and stellar age. In addition, for the first time we show that the Sun has the lowest Li abundance in comparison with solar twins at similar age (4.6 $\pm$ 0.5 Gyr). We compare the lithium content with the condensation temperature slope for a sub-sample of the best solar twins and  determine that the most lithium depleted stars also have fewer refractory elements. We speculate whether the low lithium content in the Sun might be related to the particular configuration of our Solar system.



\end{abstract}




\section{Introduction}


The importance of lithium in Astronomy ranges from cosmological to stellar evolution questions, and could even be related to exoplanets. The {\it cosmological Li problem} is related to the mismatch between the Li content produced during Big Bang nucleosynthesis and the one measured in old halo dwarf stars \citep{spite/spite/82}; a disagreement of about a factor of four is found \citep{ryan/99,asplund/06,bonifacio/07,matsuno/17}. 

In the context of Galactic chemical evolution, Li abundances obtained in thin disk stars indicate a production of this element at this component of the Galaxy, with the production mechanisms still in debate \citep{ramirez/12,bensby/18,cescutti/molaro/18,fu/18}.

Regarding stellar evolution, the Li-rich giant problem is related to how some observed giant stars have higher content of Li despite the expectation that this element is destroyed during their first dredge-up phase due to its fragile nature    \citep{casey/16,aguilera-gomez/16}, as seen in standard stellar evolution models. The work of \cite{charbonnel/10} presented a non-standard stellar evolution model considering thermohaline instability and rotational induced mixing and they were able to reproduce the Li behaviour in red giants. The  non-standard stellar nucleosynthesis presented in \cite{yan/18} might explain the observations of Li-rich giants in a  particular short stellar evolution phase.  See also recent papers by \cite{deepak/19} and \cite{casey/19}.


Despite many observational and theoretical efforts, the origin of the observed Li depletion in solar-like stars is not well established yet and remains hotly debated in the literature. Albeit likely related to internal depletion during the lifetime of the star, more Li abundances are necessary to better constrain non-standard evolution stellar models that take into consideration different internal motions of stars. The extra mixing is necessary since Li is destroyed through the reaction $^7$Li$({\rm p},\alpha)\alpha$ at temperatures of $\sim 2.5\times 10^6$ K near the base of the convective envelope in sun-like stars. Those non-standard evolution models can include gravity waves \citep{charbonnel/talon/05}, rotation-induced mixing and diffusion \citep{donascimento/09}, rotation-driven turbulent diffusion \citep{denissenkov/10}, and overshooting and gravitational settling \citep{xiong/deng/09}.


Several works in the literature discuss the factors influencing lithium depletion in solar-type stars such as  occurrence of planets \citep{mena/14}, binarity \citep{zahn/94,beck/17}, stellar age \citep{carlos/16} or even planet engulfment \citep{montalban/02,sandquist/02}. Those factors can be as important as the specific parameters of stars, which also influence the amount of lithium burning, such as the convective zone thickness that depends on the mass and metallicity of a star.


According to \citet{takeda/10} and \citet{gonzalez/10}, the difference in the stellar angular momentum could cause different amounts of Li burnt. They claim that there is an increase in the amount of Li burning, the lower the angular momentum is, thus the presence of planets or solar twins in a binary system should present different amounts of lithium in comparison with single field solar twins with the same stellar parameters and age. This is argued by \citet{israelian/09}, \citet{mena/14} and \citet{zahn/94}, but it is probably a secondary effect that accounts for only a small fraction of the total depletion \citep{pavlenko/18}. 

As discussed in \cite{beck/17}, Li abundances vary with stellar rotation  that depends on stellar age \citep{dos_santos/16}. This is in agreement with various works that indicate that the Li content in solar twins is depleted as the stars age \citep{baumann/10,monroe/13,melendeza/14,carlos/16}.  More recently, \citet{liu/16} obtained A(Li)$_{\mathrm{NLTE}}=1.36^{+0.08}_{-0.07}$ dex for one solar twin in the open cluster M67, which has a well determined age ($3.47^{+0.70}_{-0.45}$ Gyr, \citealt{gaia_col/18}). The A(Li) found is in agreement  with the relation shown in \cite{carlos/16}, based on field solar twins.


In contrast, \citet{thevenin/17} analysed solar twins with stellar parameters in the same interval as the sample of \citet{carlos/16}, and built a stellar evolution model, concluding that the Li is mainly depleted during the pre-main sequence phase, and not during all the main sequence as suggested by \citet{carlos/16}. Thus, it is imperative to our understanding of stellar interiors and the mechanisms of Li depletion to increase the solar twins sample in order to perform more detailed comparisons. 


Following the study performed by  \cite{carlos/16}, in which we analysed a sample of 21 solar twins, we present here the analysis of a broader sample of 77 solar twins plus the Sun. In this larger sample we have more than 10 new solar twins in the 0-2 Gyr age interval in contrast with just only one object at the same interval in the earlier work of \citet{carlos/16}, adding, thus, valuable information on stellar structure evolution at early ages in the main sequence phase.

The paper is organized as follows: in Section \ref{sample}, we discuss the sample and the stellar parameters adopted; in Section \ref{analysis}, we describe the analysis; Section \ref{discussion} shows the results and discussion and the conclusions are presented in Section \ref{conclusion}.

\section{SAMPLE}
\label{sample}

The sample is composed of 77 solar twins plus the Sun. The spectra are from the HARPS spectrograph \citep{mayor/03}  of the 3.6m ESO telescope at La Silla, where the solar spectrum was observed with the reflected light from the asteroid Vesta. These stars have spectra with high resolving power ($\mathrm{R}=115000$) and high signal-to-noise  ($300 \lesssim \mathrm{S/N} \lesssim 1800$).

These stars, classified as solar twins (effective temperature approximately within T$_{\mathrm{eff},\odot}\pm100$ K, surface gravity approximately within log g$_{\odot}\pm0.1$ and metallicity approximately within [Fe/H]$_{\odot}\pm0.1$), were selected from the work of \cite{ramirez/14a} and analysed in more detail by \cite{dos_santos/17}, \citet{spina/18} and \cite{bedell/18}.

The stars HIP 19911, HIP 67620 and HIP 103983 were removed from the sample due to contamination by a nearby companion, as discussed in \cite{dos_santos/17}.

In particular, \cite{spina/18} measured the stellar parameters and ages found in the present work, using high precision spectroscopy through a differential analysis (e.g., \citealt{bedell/14}). The effective temperature, surface gravity and [Fe/H] were measured using Fe I and Fe II lines in a differential line-by-line method aimed to achieve a excitation and ionisation equilibrium balance. Then, the stellar age and masses were calculated using the stellar parameters with Yonsei-Yale isochrones \citep{yi/01,kim/02}.   

In addition, the work of \cite{dos_santos/16} determined the projected rotational velocity (v$\sin i$) and macroturbulence velocity (v$_{\mathrm{macro}}$) of a bigger sample that includes all the objects studied here.

Our sample of solar twins has 17 objects in a binary system (as showed in \citealt{dos_santos/17} and references therein) and the following 5 exoplanet systems: HIP 5301 \citep{naef/10}, HIP 11915 \citep{bedell/15}, HIP 15527 \citep{jones/06}, HIP 68468 \citep{melendez/17} and HIP 116906 \citep{butler/06}. 


\section{Analysis}
\label{analysis}

The lithium abundance analysis employed here is similar to that described in \cite{carlos/16}. We applied spectral synthesis analysis in the region of the asymmetric $6707.75$ \AA\, Li I line  using the July 2014 version of the 1D LTE code MOOG \citep{sneden/73} and the Kurucz grid of ATLAS9 model atmospheres \citep{castelli/04}. As in \cite{carlos/16}, the line list from \cite{melendez/12}, which includes blends from atomic and molecular (CN and C$_2$) lines, was employed. 


In order to estimate the lithium abundances\footnote{The lithium abundances are given in the notation $A(Li) = \log(\epsilon_{Li}) = \log (N_{Li}/N_{H}) + 12$, where $N_{Li}$ and $N_{H}$ are the number densities of lithium and hydrogen respectively.} we adopted the values of T$_{\mathrm{eff}}$, [Fe/H], log g and microturbulence velocity ($\xi$) from \cite{spina/18} with v$_{\mathrm{macro}}$ and v$\sin i$ from \cite{dos_santos/16}.

\begin{figure*}

\begin{tabular}{c c}

\includegraphics[width=0.5\textwidth]{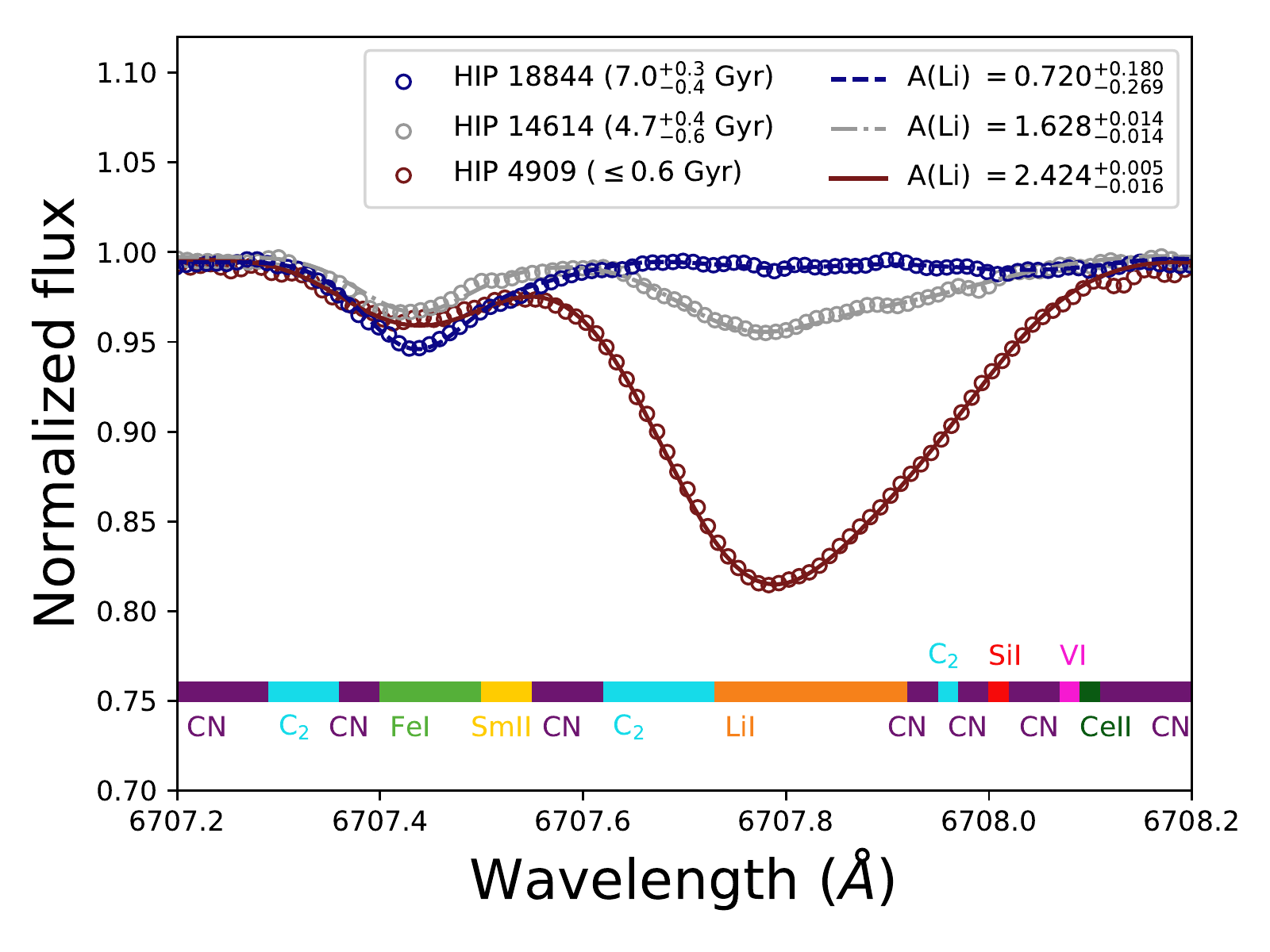} &
\includegraphics[width=0.5\textwidth]{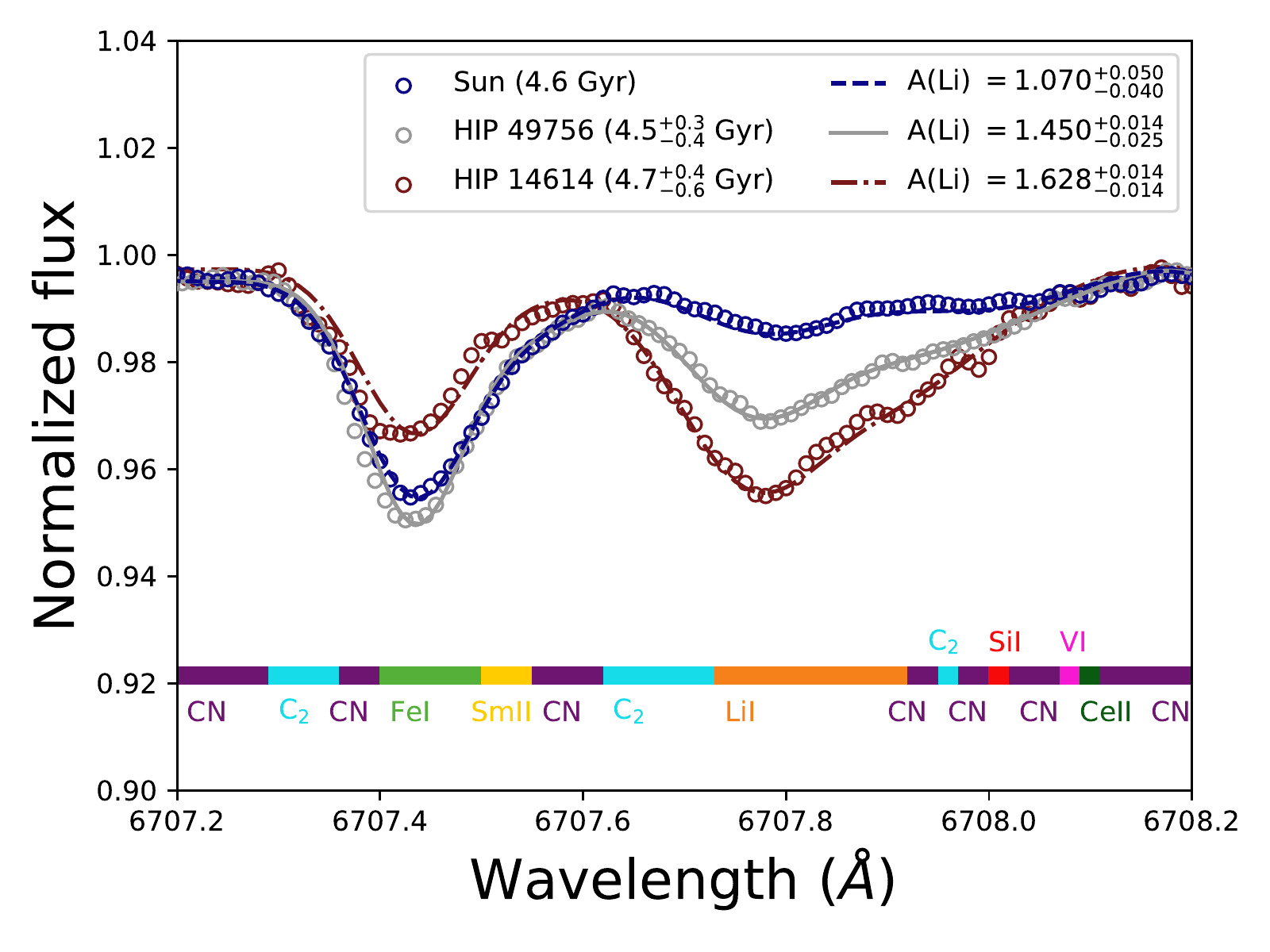}\\

\end{tabular}

\caption{Observed spectra (open circles) in comparison with their respective spectral synthesis   for three stars at different ages (left panel) and three stars at similar age (right panel). The bar at the bottom indicates the positions of each atomic or molecular species in the spectra. Notice how younger stars display higher Li abundances (left panel) and how the Sun shows a low Li abundance compared to solar twins of similar age (right panel).}
\label{obsxsint}

\end{figure*}

Figure \ref{obsxsint} shows the observed spectra in comparison with their respective spectral synthesis for different stars. It is worth noting that $^7$Li shows several components introducing an asymmetry in the profile, and the presence of $^6$Li, to a lesser extent, also contribute to this asymmetry. However, we are not considering the contribution of $^6$Li on our spectral synthesis due to its much lower abundance found in the Sun \citep{asplund/09}.

To calculate the final lithium abundance errors we considered the uncertainties in the continuum setting, the rms deviation of the observed line profile relative to the synthetic spectra, and the stellar parameters. The typical (median) Li abundance error is $\sigma=0.036$ dex.

After estimating the LTE lithium abundances, the non-LTE (NLTE) abundances were obtained through the INSPECT database\footnote{\url{www.inspect-stars.com} (version 1.0).}, based on NLTE calculations by \cite{lind/09}.  The median value of the non-LTE corrections for the whole sample is $0.04\pm0.01$ dex; the small standard deviation from the median value in comparison with the typical abundance error shows that the effect of the NLTE corrections in the differential analysis precision can be considered negligible.

The stellar parameters and  lithium abundances are presented in Table \ref{table}. We measure Li abundances down to values of about A(Li)$\sim0.6$ dex, and for the high quality spectra we achieve an upper limit of about A(Li)$\sim0.3$ dex in the most Li-poor solar twins.


\section{Discussion}
\label{discussion}

\begin{figure}

\hspace{-0.7cm}
\includegraphics[width=0.55\textwidth]{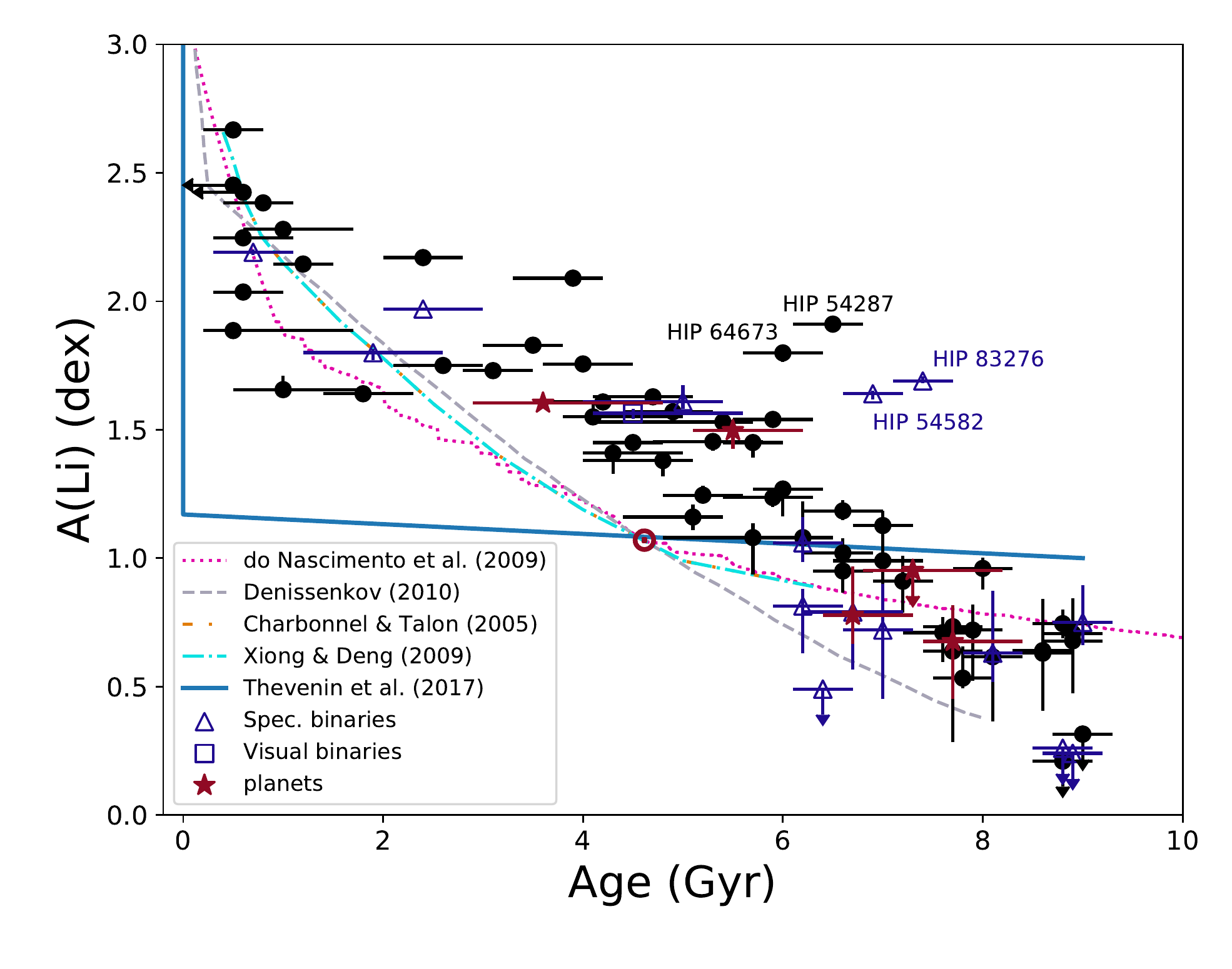}

\caption{Connection between stellar ages and NLTE lithium abundances for our sample.  
Dark red filled stars indicate stars with detected planets, dark blue squares represent visual binaries, dark blue triangles show spectral binaries and black filled circles represent single stars without planets detected. The models of Li depletion (referred in the text) were normalized to the solar Li abundance. In some cases, the lithium abundance errors are smaller than the points.}
\label{li_age_models}

\end{figure}

Figure \ref{li_age_models} shows the Li abundances versus stellar age for the whole sample. Non-standard solar models \citep{charbonnel/talon/05,donascimento/09,xiong/deng/09,denissenkov/10,thevenin/17}, calibrated to fit the Sun, are displayed for comparison. As previously discussed in \cite{carlos/16}, \cite{melendeza/14} and \cite{monroe/13}, there is a strong correlation between lithium abundances and stellar ages in solar twins (younger stars have more lithium content in comparison with older stars).

 In addition, Figure \ref{li_age_models} presents more than 10 new solar twins in the age interval $0.0\leq\mathrm{Age \, (Gyr)}\lesssim2.0$, where it is possible to notice the sharp decrease of Li abundances with age. This behaviour might be explained by the fast-rotator nature of young solar type stars \citep{pace/04,barnes/07,donascimento/14,dos_santos/16}, which may influence on internal stellar structures \citep{ballot/07,brown/08} and enhance internal transport mechanisms (e.g., \citealt{schirbel/15}), then affecting how fast Li is burnt in young solar twins. It is interesting noting that at this age interval the data is more well represented by the non-standard evolution solar model of \cite{donascimento/09}, which takes in consideration rotation induced mixing and diffusion. On the other hand, there is no agreement between data and models in the $2.0\lesssim\mathrm{Age \, (Gyr)}\lesssim4.0$ interval, where the theoretical predictions anticipate a more significant Li depletion than the shallow Li depletion observed in our sample.

\begin{figure}

\includegraphics[width=0.5\textwidth]{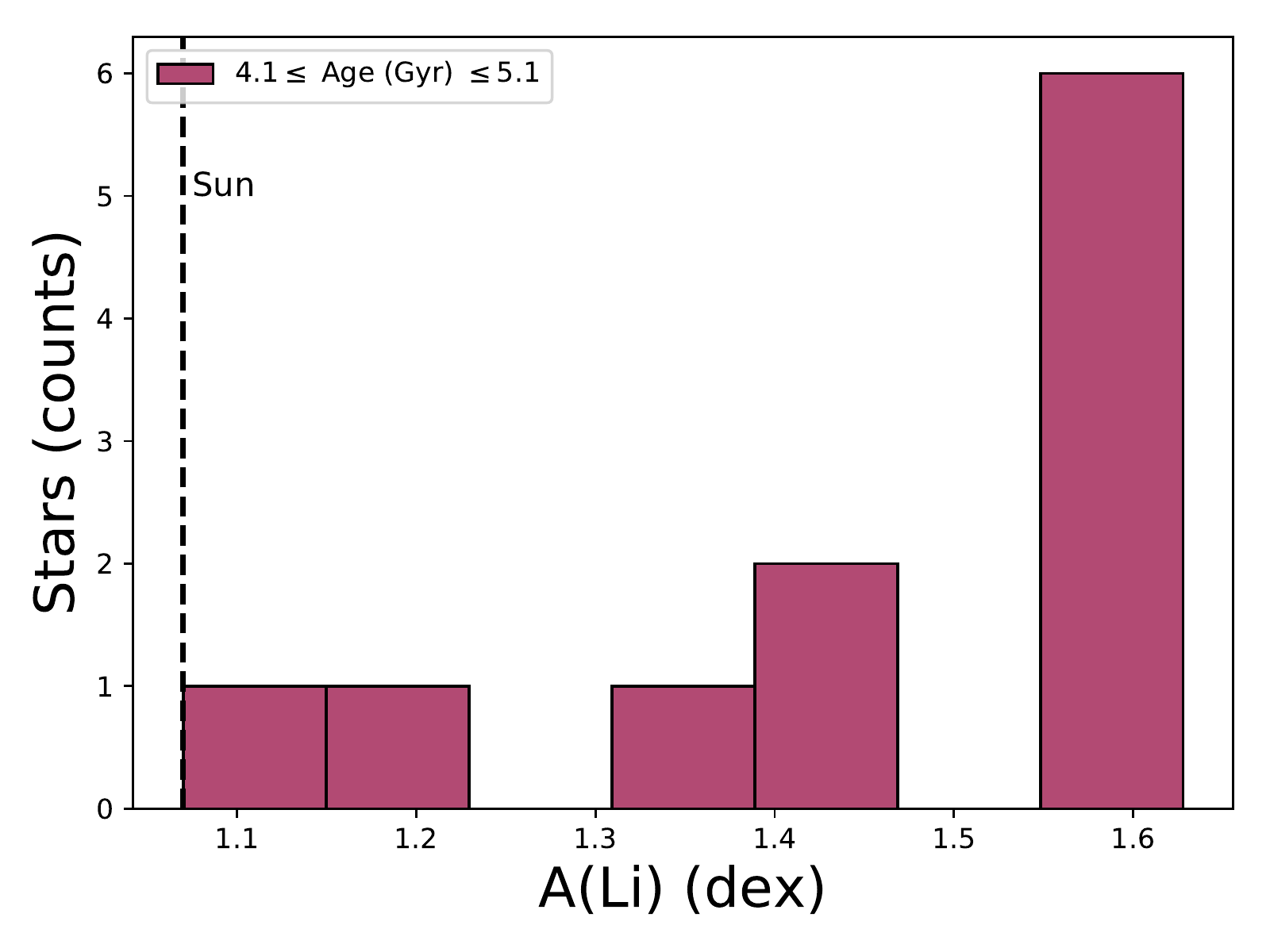}

\caption{Distribution of A(Li) in a sub-sample of solar twins with ages between 4.1 and 5.1 Gyr. The NLTE Li abundance of the Sun is shown by the black dashed line. The gaps in the distribution stress the demand of observing more solar twins at this age interval.}
\label{histog_li}

\end{figure}

Moreover, the Sun shows a lower lithium abundance in comparison with stars at same age, despite the fact that the work of \cite{dos_santos/16} and \cite{lorenzo-oliveira/18} find that the Sun has a typical rotation and activity compared with other solar twins at similar age. Figure \ref{histog_li} shows the distribution of A(Li) in solar twins at the age interval of $4.1\leq\mathrm{Age\, (Gyr)}\leq5.1$. The solar Li abundance A(Li)$_{\odot}=1.07^{+0.03}_{-0.02}$ dex is the lowest in this age interval, thus confirming that the Sun has the lowest Li abundance when compared to solar twins at similar age. Furthermore, the solar bin lies below 91\% of the sample of stars with age 4.6$\pm$0.5 Gyr.

In general, the whole sample seems to follow reasonably the A(Li) vs. age correlation with a typical scatter of $\sim 0.2$ dex at a given age, estimated from the standard deviation of the Li abundance in 1-Gyr bins (excluding the 4 outliers  HIP 54287, HIP 54582, HIP 64673 and HIP 83276, pointed in Figure \ref{li_age_models}). We perform a Spearman correlation test, considering the errors in both axes for the Li-age connection for the whole sample, excluding the outliers mentioned earlier, and find a Spearman rank coefficient r$_s=-0.95$ and a probability of $10^{-37}$ of our results arising by chance.

\begin{figure}
\begin{tabular}{c}

\includegraphics[width=0.5\textwidth]{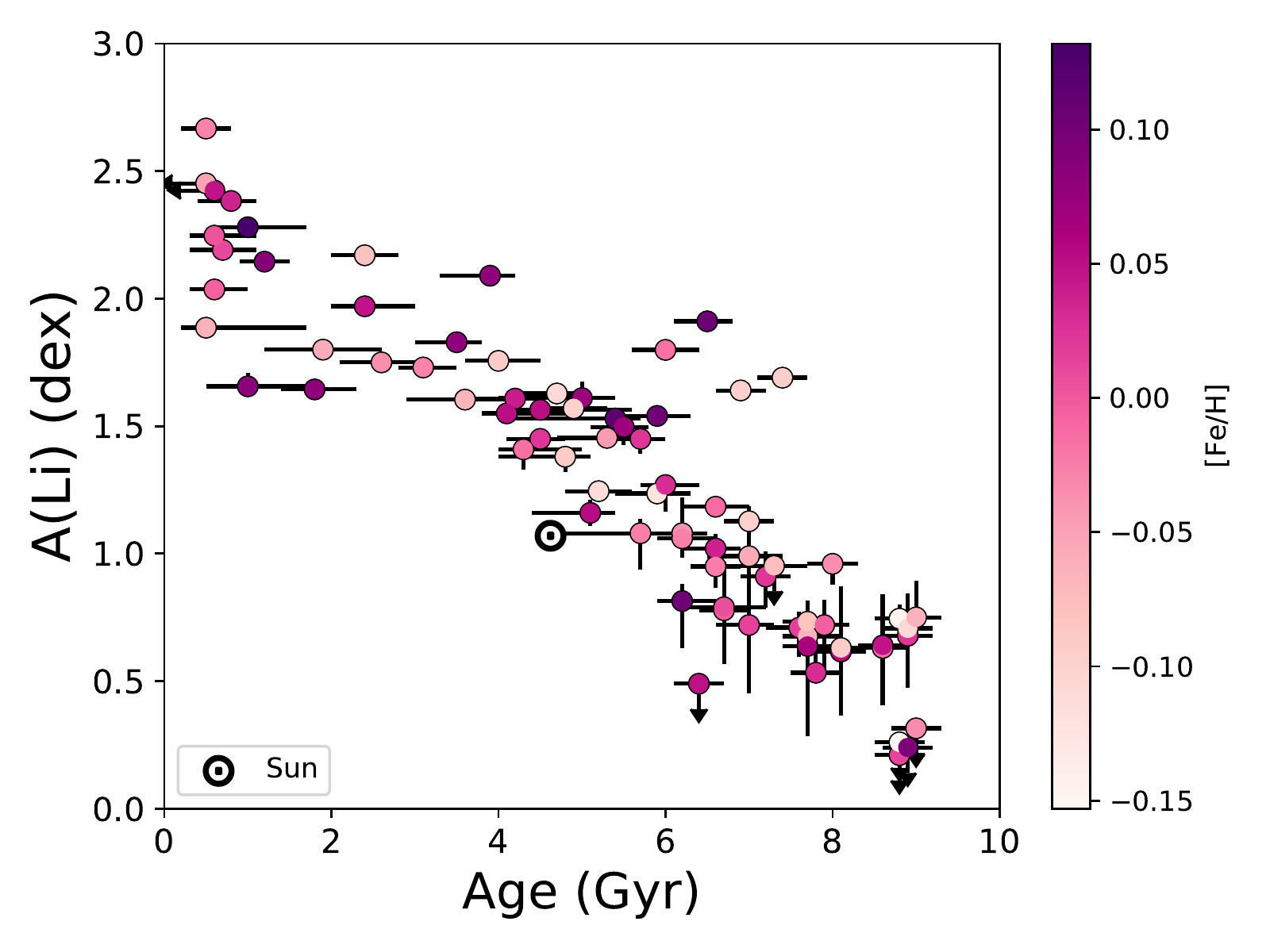} \\ \includegraphics[width=0.5\textwidth]{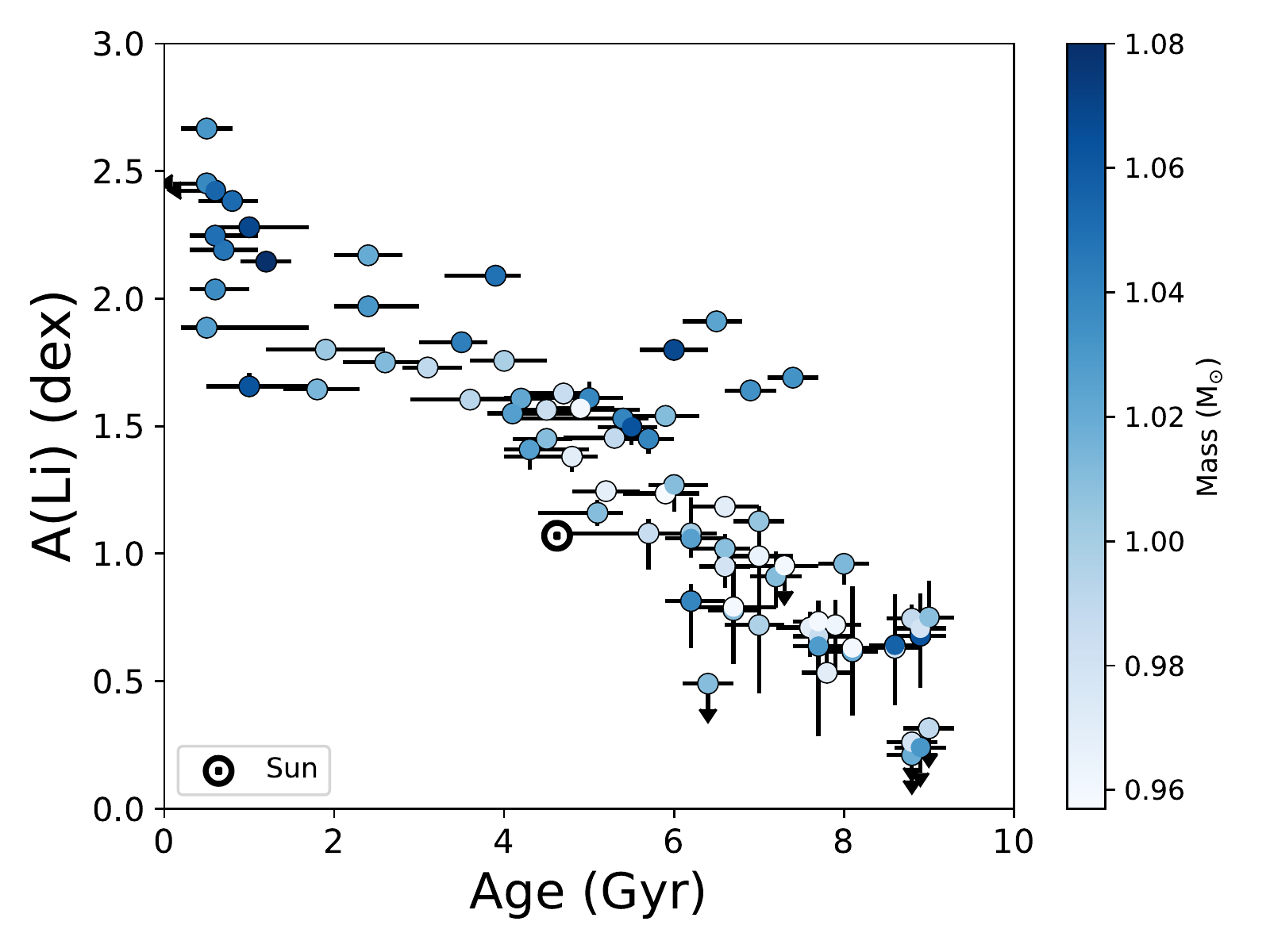} \\ \includegraphics[width=0.5\textwidth]{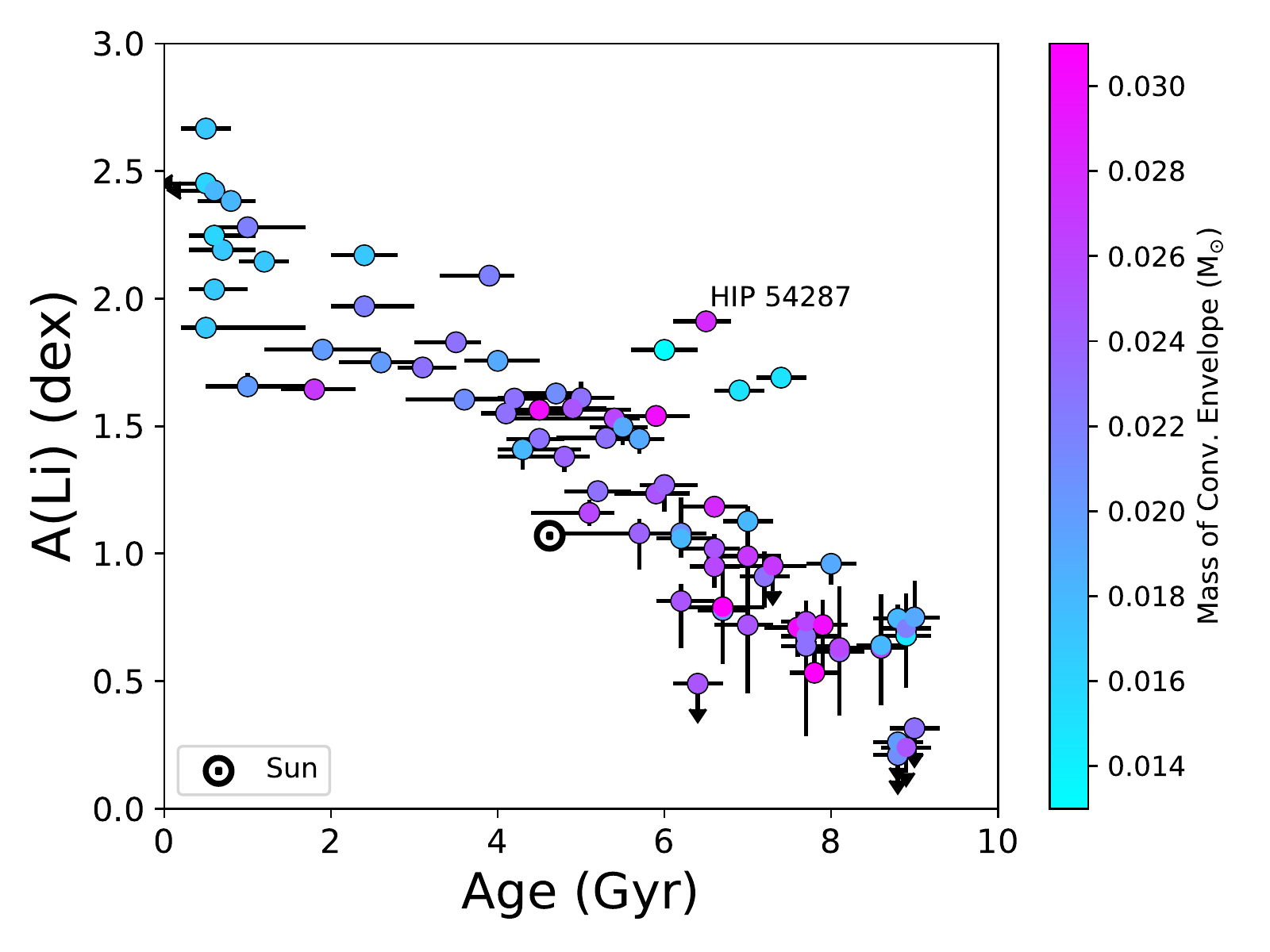}

\end{tabular}

\caption{Lithium abundances versus stellar age colour coded by [Fe/H] (top panel), mass (middle panel), and the mass of the convective envelope (bottom panel). HIP 54287 is labelled in the lower panel because, as discussed in the text, it could have engulfed a planet.}
\label{cm_feh_mass_cev}

\end{figure}

To shed some light on these outliers we analyse separately the correlation between A(Li) versus stellar age with [Fe/H], stellar mass and the mass of the convective envelope (Figure \ref{cm_feh_mass_cev}), where the mass of the convective envelope was calculated by interpolating the values found in the YaPSI\footnote{Yale Astro web page: \protect\url{http://www.astro.yale.edu/yapsi/}; AIP web page: \protect\url{http://vo.aip.de/yapsi/}.} grid of isochrones \citep{spada/16}.

The upper panel of Figure \ref{cm_feh_mass_cev} shows the dependence between A(Li) and stellar age with [Fe/H]  (the typical error is $\sigma($[Fe/H]$)=0.004$, \citealt{spina/18}). We conclude that the sample is homogeneous regarding metallicity and stellar age for this interval; and due to the fact that our sample is composed by only solar twins ($-0.1\lesssim$[Fe/H]$\lesssim0.1$), there is no apparent trend in Li abundances with [Fe/H] for a given age. In addition, the  outliers HIP 54287, HIP 54582, HIP 64673 and HIP 83276 have substantial differences in [Fe/H] varying from $-0.096$ dex to $0.107$ dex. 


In the middle panel of Figure \ref{cm_feh_mass_cev} we present the correlation between A(Li), stellar age and masses (the typical error is $\sigma(\mathrm{M/M}_{\odot})=0.004$, \citealt{spina/18}). Likewise the [Fe/H], the stellar mass distribution is somewhat homogeneous in all the age interval,  apart from the youngest stars with age $\lesssim 2.0$ Gyr where we lack stars with mass $\lesssim0.98 \mathrm{M}_{\odot}$.

The lower panel of Figure \ref{cm_feh_mass_cev} displays A(Li), stellar ages and masses of the convective envelopes for the whole sample. Following the dependence in mass and [Fe/H] shown in the upper and middle panel of Figure \ref{cm_feh_mass_cev}, the sample is somewhat homogeneous for stars with age $\gtrsim 2$ Gyr without considering the outliers mentioned earlier. 

Discussing specifically the outliers in our sample, three of the four objects (HIP 54582, HIP 64673 and HIP 83276) present a less massive convective envelope; as seen in the lower panel of Figure \ref{cm_feh_mass_cev}. This might be an effect of the combination of the lower values of [Fe/H] and higher values of stellar mass in comparison with other objects in the sample. Thus,  the small size of the convective envelope implies in less Li burning, which causes the discrepancy in the Li content in these three stars in comparison to the rest of the sample.  

Although the star HIP 54287 has a ``regular'' convective envelope  to burn Li at the same extent as  other stars at the same bin of age (excluding the outliers), the high Li content indicates that this object could have experienced a planet engulfment, as described in \cite{montalban/02} and \cite{sandquist/02} and as previously discussed in \cite{carlos/16}. If this is the case, we are observing a short duration event, because according to \cite{theado/vauclair/12}, thermohaline mixing should dilute the Li overabundance that we observe in about $\sim50$ million years, or perhaps thermohaline mixing could be less efficient and the Li enhancement remain for longer times (increasing the probability of observing this type of event).

\begin{figure}

\includegraphics[width=0.5\textwidth]{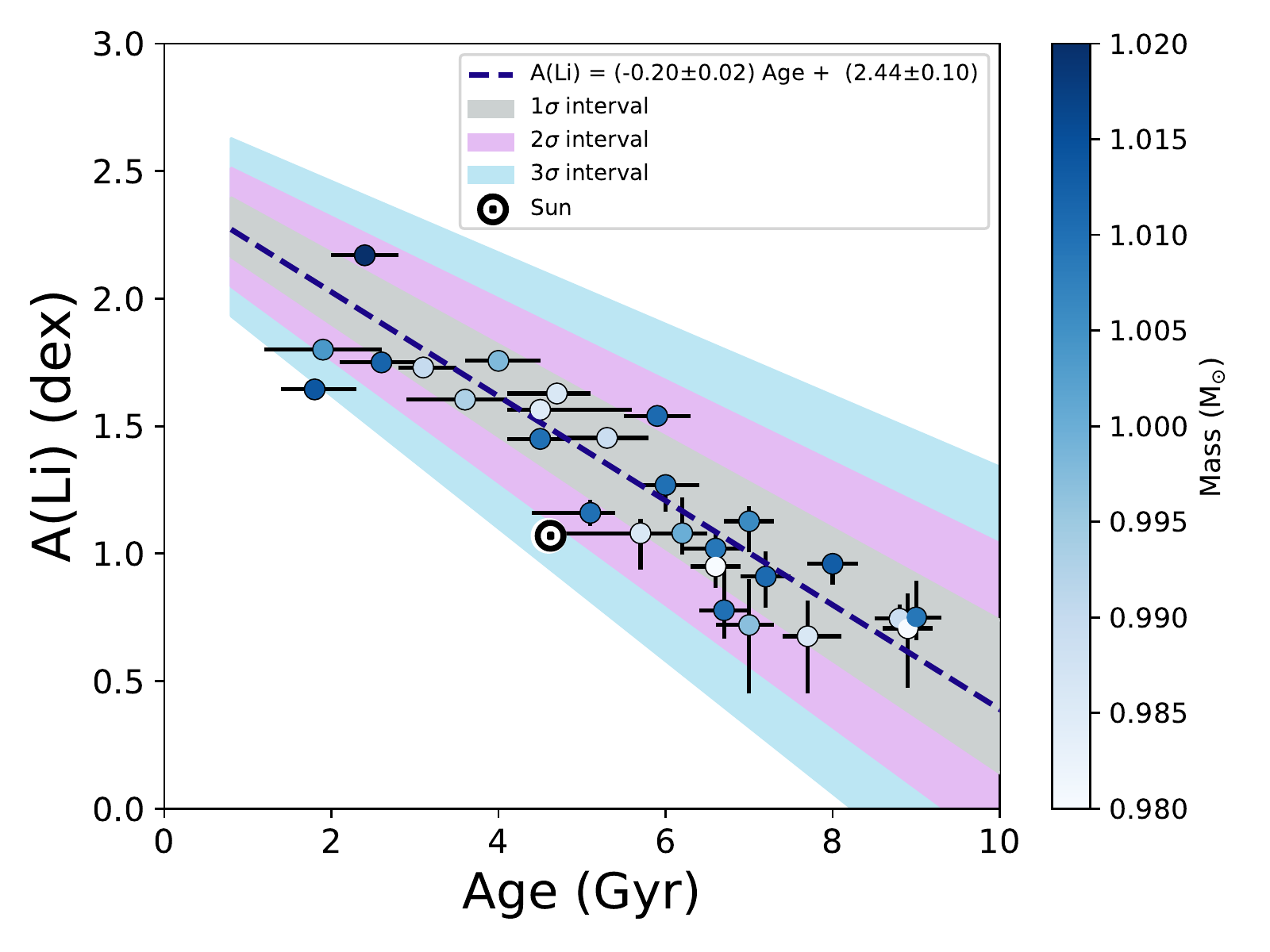}

\caption{Lithium abundances versus stellar age as a function of mass in the interval $0.98\leq\mathrm{M}/\mathrm{M}_{\odot}\leq1.02$.}
\label{li_age_cmmass_mascara}

\end{figure}

The data presented in Figure \ref{li_age_cmmass_mascara} is for the case when  we narrow our criteria of solar twins for our sample and consider only objects with mass in the interval 0.98 $ \leq \mathrm{M/M} _{\odot} \leq $ 1.02, and excluding upper limits in Li abundances.  For this sub-sample, the median value for the masses of the convective envelopes is $0.023\pm0.003$ dex, confirming their similarity.

In this case, the best linear fit, calculated using orthogonal distance regression with the {\it scipy.odr\footnote{\url{http://docs.scipy.org/doc/scipy/
reference/odr.html}}} (hereafter ODR) considering the errors in both axes, between lithium abundances and stellar ages is:

\begin{equation}
\mathrm{A(Li)} = (-0.20\pm0.02)\mathrm{Age} + (2.44\pm0.10)
\label{eq1}
\end{equation}

We can notice in Figure \ref{li_age_cmmass_mascara} how the Sun is Li-poor when compared to solar twins at the same age  (see also the right panel of Figure \ref{obsxsint} and Figure \ref{histog_li}). The Sun is an outlier by $\sim2\sigma$ from the Li-age correlation given in Equation \ref{eq1}; furthermore, the Sun has the lowest Li abundance among solar twins of similar age (4.6 Gyr).

Using the Equation \ref{eq1} we found,   for the whole sample, a correspondence between the lithium residuals ($\Delta$A(Li)) and the stellar parameters, mass and [Fe/H]: 




\begin{equation}
\begin{split}
\Delta\mathrm{A(Li)} = \,\,&- (3.55\pm1.11) + (3.47\pm1.09)\mathrm{M/M}_{\odot}\\
                       &- (1.17\pm0.52)\mathrm{[Fe/H]}
\end{split}
\end{equation}

This equation is in concordance with the models of \cite{castro/09}; who find that the lithium burning increases with increasing [Fe/H]; the opposite occurs for the other parameter where the lithium depletion is accentuated with decreasing in stellar mass. Furthermore, this result is compatible with the respective equation from \cite{carlos/16} and have a more significant dependence in comparison with the one presented in this earlier work (3.2$\sigma$ against 1.6$\sigma$ for stellar mass and 2.3$\sigma$ against 1.7$\sigma$ for [Fe/H]).



\begin{figure}

\includegraphics[width=0.5\textwidth]{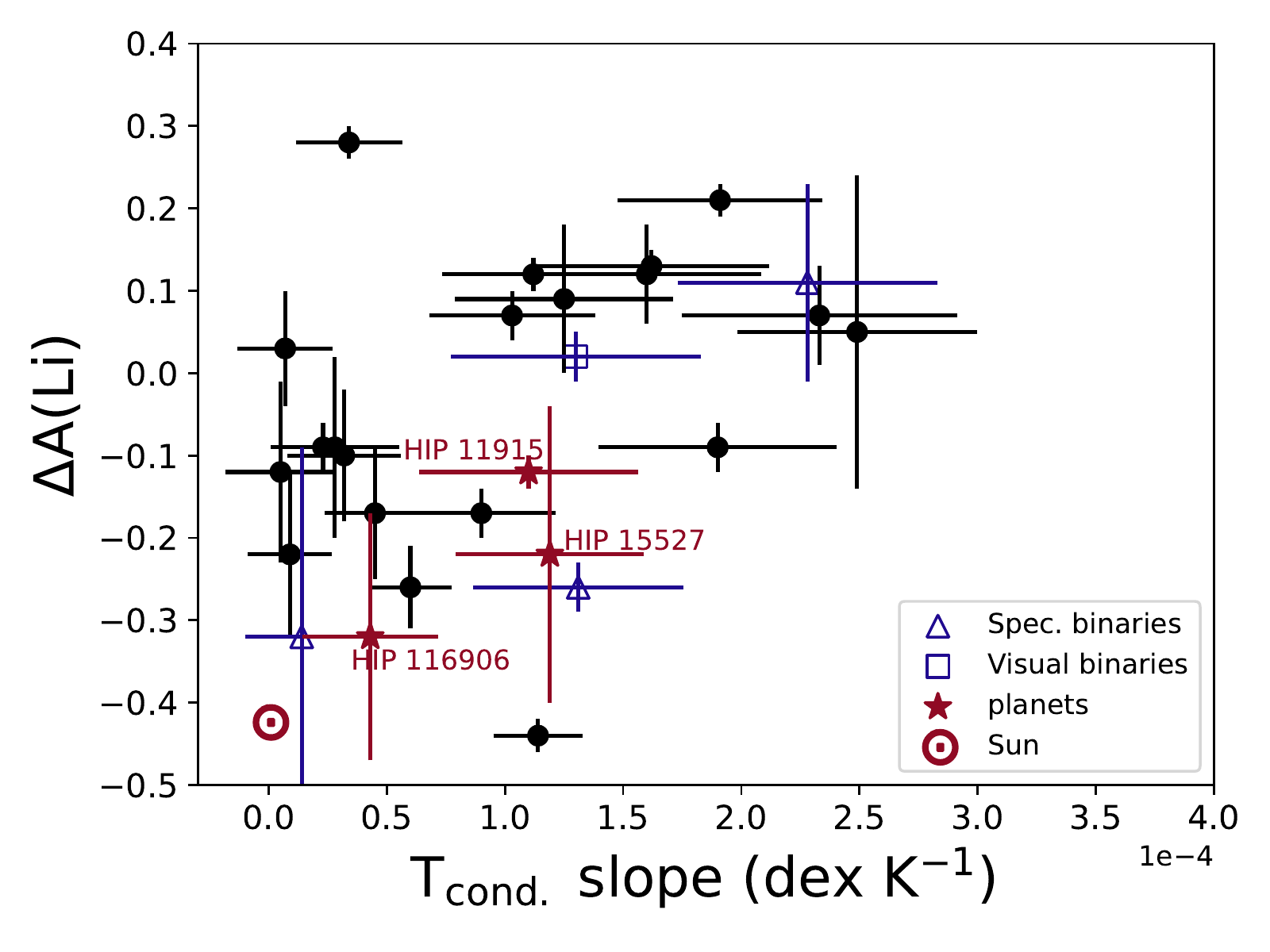}

\caption{The lithium residuals (from Figure \protect\ref{li_age_cmmass_mascara}) against the condensation temperature  (T$_{\mathrm{cond.}}$) slope from \protect\cite{bedell/18} for our subsample of stars with 0.98 $ \leq \mathrm{M/M} _{\odot} \leq $ 1.02. The name of the stars with detected planets are shown.}
\label{slope_tc}

\end{figure}

Figure \ref{slope_tc} shows $\Delta$A(Li) for our subsample of stars with 0.98 $ \leq \mathrm{M/M} _{\odot} \leq $ 1.02 versus their respective condensation temperature  slope  corrected by the galactic chemical evolution (T$_\mathrm{c}$ slope, third column of Table 4  from \citealt{bedell/18}).  

The analysis of the condensation temperature trends can shed some light on the planetary formation scenarios since these T$_{\mathrm{c}}$ slopes are linked with the content of refractory elements (T $\gtrsim 900$ K, see discussion in \citealt{bedell/18}) in stellar atmospheres, which can be associated to  rocky planets \citep{melendez/09}.


The three stars with planets detected presented in Figure \ref{slope_tc} are: HIP 11915 that has a Jupiter twin detected with orbital period of $3830\pm150$ days and $\mathrm{m}_\mathrm{p}sini=0.99\pm0.06$ M$_{\mathrm{jup}}$ \citep{bedell/15}; HIP 116906 with one planet detected with orbital period of $572.38\pm0.61$ days and $\mathrm{m}_\mathrm{p}sini=7.75\pm0.65$ M$_{\mathrm{jup}}$ \citep{butler/06}; and HIP 15527 with one planet detected with orbital period of $595.86\pm0.03$ days and $\mathrm{m}_\mathrm{p}sini=1.77\pm0.22$ M$_{\mathrm{jup}}$ \citep{jones/06}.

We perform a Spearman correlation test considering the errors in both axes for the sample shown in Figure \ref{slope_tc}  and found a Spearman rank coefficient $\mathrm{r}_s = 0.47$ and a probability of $0.01$ of our results arising by chance. 






This tentative correlation indicates that the more lithium depleted stars have less content of refractory material. According to \cite{bedell/18}, the Sun presents a refractory-to-volatile deficiency relative to 93\% of the sample of solar twins. If this depletion in refractory material is connected to the presence of planets or even the quantity of planets in our planetary system, what Figure \ref{slope_tc} shows is that the low lithium content of the Sun might be linked to the presence of rocky planets.  Although we should be cautious as we do not have a complete census down to Earth masses of the planets hosted by other solar twins, the lower solar lithium  content in comparison with stars at the same bin of age could be related to our solar system configuration and possibly to terrestrial planets.


In addition, according to \cite{tucci/15} the Sun is also poor in Be. Despite the shallow trend of Be content with age for solar twins, \cite{tucci/15} showed that the solar Be is less abundant by $\sim 0.05$ dex in comparison with other solar twins. This deficiency in Be might be linked to the Sun lower content of refractory material as well, as pointed out in \cite{tucci/15}. Interestingly, the work of \cite{botelho/19} found that the Sun has a lower [Th/Fe] ratio compared to solar twins at similar age, and also when corrected to its ZAMS (zero age main sequence) value. This somewhat lower abundance of Th in the Sun is perhaps because Th is a highly refractory element \citep{lodders/03}, reflecting thus the refractory-depleted composition of the Sun, which could be linked to rocky planets \citep{melendez/09}.

\section{Conclusions}
\label{conclusion}

 We measured high-precision lithium abundances (median error of $0.036$ dex) for a sample of 77 solar twins with high resolution and high signal to noise spectra from the HARPS spectrograph.

We confirm previous results showing the strong connection between lithium depletion and stellar ages  and also identified a steeper Li depletion with stellar age for young solar twins (Age $\lesssim2.0$ Gyr). 

Three of the four outliers in this work can be explained when considering the respective masses of their convective envelopes.

 It seems that there is no significant difference in lithium depletion between  known planet host stars and stars with no planets detected, when we analyse the lithium depletion and stellar age correlation. The same behaviour is found for visual and spectral binaries in comparison with single field stars.

 We found that the Sun can be considered a lithium-poor star in comparison with other solar twins at similar age (by a factor of $\sim2\sigma$). Also, our data suggest that stars with the lowest Li abundances are accompanied by a lower level of refractory elements. This could be explained by the presence of rocky planets and the unique architecture of the solar system.




\section*{Acknowledgements}

MC would like to acknowledge support from CAPES. This study was financed in part by the Coordena\c{c}\~ao de Aperfei\c{c}oamento de Pessoal de N\'ivel Superior - Brasil (CAPES) - Finance Code 001. JM is thankful for the support of FAPESP (2014/18100-4, 2018/04055-8) and CNPq (Bolsa de Produtividade).  LAdS acknowledges the financial support from the European Research Council (ERC) under the European Union's Horizon 2020 research and innovation program (project {\sc Four Aces}; grant agreement No. 724427).




\bibliographystyle{mnras}
\bibliography{mybib} 

\begin{thebibliography}{}
\makeatletter
\relax
\def\mn@urlcharsother{\let\do\@makeother \do\$\do\&\do\#\do\^\do\_\do\%\do\~}
\def\mn@doi{\begingroup\mn@urlcharsother \@ifnextchar [ {\mn@doi@}
  {\mn@doi@[]}}
\def\mn@doi@[#1]#2{\def\@tempa{#1}\ifx\@tempa\@empty \href
  {http://dx.doi.org/#2} {doi:#2}\else \href {http://dx.doi.org/#2} {#1}\fi
  \endgroup}
\def\mn@eprint#1#2{\mn@eprint@#1:#2::\@nil}
\def\mn@eprint@arXiv#1{\href {http://arxiv.org/abs/#1} {{\tt arXiv:#1}}}
\def\mn@eprint@dblp#1{\href {http://dblp.uni-trier.de/rec/bibtex/#1.xml}
  {dblp:#1}}
\def\mn@eprint@#1:#2:#3:#4\@nil{\def\@tempa {#1}\def\@tempb {#2}\def\@tempc
  {#3}\ifx \@tempc \@empty \let \@tempc \@tempb \let \@tempb \@tempa \fi \ifx
  \@tempb \@empty \def\@tempb {arXiv}\fi \@ifundefined
  {mn@eprint@\@tempb}{\@tempb:\@tempc}{\expandafter \expandafter \csname
  mn@eprint@\@tempb\endcsname \expandafter{\@tempc}}}

\bibitem[\protect\citeauthoryear{{Aguilera-G{\'o}mez}, {Chanam{\'e}},
  {Pinsonneault}  \& {Carlberg}}{{Aguilera-G{\'o}mez}
  et~al.}{2016}]{aguilera-gomez/16}
{Aguilera-G{\'o}mez} C.,  {Chanam{\'e}} J.,  {Pinsonneault} M.~H.,   {Carlberg}
  J.~K.,  2016, \mn@doi [\apj] {10.3847/0004-637X/829/2/127}, \href
  {http://adsabs.harvard.edu/abs/2016ApJ...829..127A} {829, 127}

\bibitem[\protect\citeauthoryear{{Asplund}, {Lambert}, {Nissen}, {Primas}  \&
  {Smith}}{{Asplund} et~al.}{2006}]{asplund/06}
{Asplund} M.,  {Lambert} D.~L.,  {Nissen} P.~E.,  {Primas} F.,   {Smith} V.~V.,
   2006, \mn@doi [\apj] {10.1086/503538}, \href
  {http://adsabs.harvard.edu/abs/2006ApJ...644..229A} {644, 229}

\bibitem[\protect\citeauthoryear{{Asplund}, {Grevesse}, {Sauval}  \&
  {Scott}}{{Asplund} et~al.}{2009}]{asplund/09}
{Asplund} M.,  {Grevesse} N.,  {Sauval} A.~J.,   {Scott} P.,  2009, \mn@doi
  [\araa] {10.1146/annurev.astro.46.060407.145222}, \href
  {http://adsabs.harvard.edu/abs/2009ARA%26A..47..481A} {47, 481}

\bibitem[\protect\citeauthoryear{{Ballot}, {Brun}  \&
  {Turck-Chi{\`e}ze}}{{Ballot} et~al.}{2007}]{ballot/07}
{Ballot} J.,  {Brun} A.~S.,   {Turck-Chi{\`e}ze} S.,  2007, \mn@doi [\apj]
  {10.1086/521617}, \href {http://adsabs.harvard.edu/abs/2007ApJ...669.1190B}
  {669, 1190}

\bibitem[\protect\citeauthoryear{{Barnes}}{{Barnes}}{2007}]{barnes/07}
{Barnes} S.~A.,  2007, \mn@doi [\apj] {10.1086/519295}, \href
  {http://adsabs.harvard.edu/abs/2007ApJ...669.1167B} {669, 1167}

\bibitem[\protect\citeauthoryear{{Baumann}, {Ram{\'{\i}}rez}, {Mel{\'e}ndez},
  {Asplund}  \& {Lind}}{{Baumann} et~al.}{2010}]{baumann/10}
{Baumann} P.,  {Ram{\'{\i}}rez} I.,  {Mel{\'e}ndez} J.,  {Asplund} M.,   {Lind}
  K.,  2010, \mn@doi [\aap] {10.1051/0004-6361/201015137}, \href
  {http://adsabs.harvard.edu/abs/2010A%26A...519A..87B} {519, A87}

\bibitem[\protect\citeauthoryear{{Beck} et~al.,}{{Beck} et~al.}{2017}]{beck/17}
{Beck} P.~G.,  et~al., 2017, \mn@doi [\aap] {10.1051/0004-6361/201629820},
  \href {http://adsabs.harvard.edu/abs/2017A%26A...602A..63B} {602, A63}

\bibitem[\protect\citeauthoryear{{Bedell}, {Mel{\'e}ndez}, {Bean},
  {Ram{\'{\i}}rez}, {Leite}  \& {Asplund}}{{Bedell} et~al.}{2014}]{bedell/14}
{Bedell} M.,  {Mel{\'e}ndez} J.,  {Bean} J.~L.,  {Ram{\'{\i}}rez} I.,  {Leite}
  P.,   {Asplund} M.,  2014, \mn@doi [\apj] {10.1088/0004-637X/795/1/23}, \href
  {http://adsabs.harvard.edu/abs/2014ApJ...795...23B} {795, 23}

\bibitem[\protect\citeauthoryear{{Bedell} et~al.,}{{Bedell}
  et~al.}{2015}]{bedell/15}
{Bedell} M.,  et~al., 2015, \mn@doi [\aap] {10.1051/0004-6361/201525748}, \href
  {https://ui.adsabs.harvard.edu/#abs/2015A&A...581A..34B} {581, A34}

\bibitem[\protect\citeauthoryear{{Bedell} et~al.,}{{Bedell}
  et~al.}{2018}]{bedell/18}
{Bedell} M.,  et~al., 2018, \mn@doi [\apj] {10.3847/1538-4357/aad908}, \href
  {http://adsabs.harvard.edu/abs/2018ApJ...865...68B} {865, 68}

\bibitem[\protect\citeauthoryear{{Bensby} \& {Lind}}{{Bensby} \&
  {Lind}}{2018}]{bensby/18}
{Bensby} T.,  {Lind} K.,  2018, \mn@doi [\aap] {10.1051/0004-6361/201833118},
  \href {http://adsabs.harvard.edu/abs/2018A%26A...615A.151B} {615, A151}

\bibitem[\protect\citeauthoryear{{Bonifacio} et~al.,}{{Bonifacio}
  et~al.}{2007}]{bonifacio/07}
{Bonifacio} P.,  et~al., 2007, \mn@doi [\aap] {10.1051/0004-6361:20064834},
  \href {http://adsabs.harvard.edu/abs/2007A%26A...462..851B} {462, 851}

\bibitem[\protect\citeauthoryear{{Botelho} et~al.,}{{Botelho}
  et~al.}{2019}]{botelho/19}
{Botelho} R.~B.,  et~al., 2019, \mn@doi [\mnras] {10.1093/mnras/sty2791}, \href
  {http://adsabs.harvard.edu/abs/2019MNRAS.482.1690B} {482, 1690}

\bibitem[\protect\citeauthoryear{{Brown}, {Browning}, {Brun}, {Miesch}  \&
  {Toomre}}{{Brown} et~al.}{2008}]{brown/08}
{Brown} B.~P.,  {Browning} M.~K.,  {Brun} A.~S.,  {Miesch} M.~S.,   {Toomre}
  J.,  2008, \mn@doi [\apj] {10.1086/592397}, \href
  {http://adsabs.harvard.edu/abs/2008ApJ...689.1354B} {689, 1354}

\bibitem[\protect\citeauthoryear{{Butler} et~al.,}{{Butler}
  et~al.}{2006}]{butler/06}
{Butler} R.~P.,  et~al., 2006, \mn@doi [\apj] {10.1086/504701}, \href
  {http://adsabs.harvard.edu/abs/2006ApJ...646..505B} {646, 505}

\bibitem[\protect\citeauthoryear{{Carlos}, {Nissen}  \&
  {Mel{\'e}ndez}}{{Carlos} et~al.}{2016}]{carlos/16}
{Carlos} M.,  {Nissen} P.~E.,   {Mel{\'e}ndez} J.,  2016, \mn@doi [\aap]
  {10.1051/0004-6361/201527478}, \href
  {http://adsabs.harvard.edu/abs/2016A%26A...587A.100C} {587, A100}

\bibitem[\protect\citeauthoryear{{Casey} et~al.,}{{Casey}
  et~al.}{2016}]{casey/16}
{Casey} A.~R.,  et~al., 2016, \mn@doi [\mnras] {10.1093/mnras/stw1512}, \href
  {http://adsabs.harvard.edu/abs/2016MNRAS.461.3336C} {461, 3336}

\bibitem[\protect\citeauthoryear{{Casey} et~al.,}{{Casey}
  et~al.}{2019}]{casey/19}
{Casey} A.~R.,  et~al., 2019, arXiv e-prints, \href
  {http://adsabs.harvard.edu/abs/2019arXiv190204102C} {}

\bibitem[\protect\citeauthoryear{{Castelli} \& {Kurucz}}{{Castelli} \&
  {Kurucz}}{2004}]{castelli/04}
{Castelli} F.,  {Kurucz} R.~L.,  2004, ArXiv Astrophysics e-prints, \href
  {http://adsabs.harvard.edu/abs/2004astro.ph..5087C} {}

\bibitem[\protect\citeauthoryear{{Castro}, {Vauclair}, {Richard}  \&
  {Santos}}{{Castro} et~al.}{2009}]{castro/09}
{Castro} M.,  {Vauclair} S.,  {Richard} O.,   {Santos} N.~C.,  2009, \mn@doi
  [\aap] {10.1051/0004-6361:20078928}, \href
  {http://adsabs.harvard.edu/abs/2009A%26A...494..663C} {494, 663}

\bibitem[\protect\citeauthoryear{{Cescutti} \& {Molaro}}{{Cescutti} \&
  {Molaro}}{2018}]{cescutti/molaro/18}
{Cescutti} G.,  {Molaro} P.,  2018, preprint, \href
  {http://adsabs.harvard.edu/abs/2018arXiv181012215C} {} (\mn@eprint {arXiv}
  {1810.12215})

\bibitem[\protect\citeauthoryear{{Charbonnel} \& {Lagarde}}{{Charbonnel} \&
  {Lagarde}}{2010}]{charbonnel/10}
{Charbonnel} C.,  {Lagarde} N.,  2010, \mn@doi [\aap]
  {10.1051/0004-6361/201014432}, \href
  {http://adsabs.harvard.edu/abs/2010A%26A...522A..10C} {522, A10}

\bibitem[\protect\citeauthoryear{{Charbonnel} \& {Talon}}{{Charbonnel} \&
  {Talon}}{2005}]{charbonnel/talon/05}
{Charbonnel} C.,  {Talon} S.,  2005, \mn@doi [Science]
  {10.1126/science.1116849}, \href
  {http://adsabs.harvard.edu/abs/2005Sci...309.2189C} {309, 2189}

\bibitem[\protect\citeauthoryear{{Deepak} \& {Reddy}}{{Deepak} \&
  {Reddy}}{2019}]{deepak/19}
{Deepak} {Reddy} B.~E.,  2019, \mn@doi [\mnras] {10.1093/mnras/stz128}, \href
  {http://adsabs.harvard.edu/abs/2019MNRAS.484.2000D} {484, 2000}

\bibitem[\protect\citeauthoryear{{Delgado Mena} et~al.,}{{Delgado Mena}
  et~al.}{2014}]{mena/14}
{Delgado Mena} E.,  et~al., 2014, \mn@doi [\aap] {10.1051/0004-6361/201321493},
  \href {http://adsabs.harvard.edu/abs/2014A%26A...562A..92D} {562, A92}

\bibitem[\protect\citeauthoryear{{Denissenkov}}{{Denissenkov}}{2010}]{denissenkov/10}
{Denissenkov} P.~A.,  2010, \mn@doi [\apj] {10.1088/0004-637X/719/1/28}, \href
  {http://adsabs.harvard.edu/abs/2010ApJ...719...28D} {719, 28}

\bibitem[\protect\citeauthoryear{{Do Nascimento}, {Castro}, {Mel{\'e}ndez},
  {Bazot}, {Th{\'e}ado}, {Porto de Mello}  \& {de Medeiros}}{{Do Nascimento}
  et~al.}{2009}]{donascimento/09}
{Do Nascimento} Jr. J.~D.,  {Castro} M.,  {Mel{\'e}ndez} J.,  {Bazot} M.,
  {Th{\'e}ado} S.,  {Porto de Mello} G.~F.,   {de Medeiros} J.~R.,  2009,
  \mn@doi [\aap] {10.1051/0004-6361/200911935}, \href
  {http://adsabs.harvard.edu/abs/2009A%26A...501..687D} {501, 687}

\bibitem[\protect\citeauthoryear{{Fu} et~al.,}{{Fu} et~al.}{2018}]{fu/18}
{Fu} X.,  et~al., 2018, \mn@doi [\aap] {10.1051/0004-6361/201731677}, \href
  {http://adsabs.harvard.edu/abs/2018A%26A...610A..38F} {610, A38}

\bibitem[\protect\citeauthoryear{{Gaia Collaboration} et~al.,}{{Gaia
  Collaboration} et~al.}{2018}]{gaia_col/18}
{Gaia Collaboration} et~al., 2018, \mn@doi [\aap]
  {10.1051/0004-6361/201832843}, \href
  {https://ui.adsabs.harvard.edu/#abs/2018A&A...616A..10G} {616, A10}

\bibitem[\protect\citeauthoryear{{Gonzalez}, {Carlson}  \& {Tobin}}{{Gonzalez}
  et~al.}{2010}]{gonzalez/10}
{Gonzalez} G.,  {Carlson} M.~K.,   {Tobin} R.~W.,  2010, \mn@doi [\mnras]
  {10.1111/j.1365-2966.2009.16195.x}, \href
  {http://adsabs.harvard.edu/abs/2010MNRAS.403.1368G} {403, 1368}

\bibitem[\protect\citeauthoryear{{Israelian} et~al.,}{{Israelian}
  et~al.}{2009}]{israelian/09}
{Israelian} G.,  et~al., 2009, \mn@doi [\nat] {10.1038/nature08483}, \href
  {http://adsabs.harvard.edu/abs/2009Natur.462..189I} {462, 189}

\bibitem[\protect\citeauthoryear{{Jones}, {Butler}, {Tinney}, {Marcy},
  {Carter}, {Penny}, {McCarthy}  \& {Bailey}}{{Jones} et~al.}{2006}]{jones/06}
{Jones} H.~R.~A.,  {Butler} R.~P.,  {Tinney} C.~G.,  {Marcy} G.~W.,  {Carter}
  B.~D.,  {Penny} A.~J.,  {McCarthy} C.,   {Bailey} J.,  2006, \mn@doi [\mnras]
  {10.1111/j.1365-2966.2006.10298.x}, \href
  {http://adsabs.harvard.edu/abs/2006MNRAS.369..249J} {369, 249}

\bibitem[\protect\citeauthoryear{{Kim}, {Demarque}, {Yi}  \& {Alexander}}{{Kim}
  et~al.}{2002}]{kim/02}
{Kim} Y.-C.,  {Demarque} P.,  {Yi} S.~K.,   {Alexander} D.~R.,  2002, \mn@doi
  [\apjs] {10.1086/343041}, \href
  {http://adsabs.harvard.edu/abs/2002ApJS..143..499K} {143, 499}

\bibitem[\protect\citeauthoryear{{Lind}, {Asplund}  \& {Barklem}}{{Lind}
  et~al.}{2009}]{lind/09}
{Lind} K.,  {Asplund} M.,   {Barklem} P.~S.,  2009, \mn@doi [\aap]
  {10.1051/0004-6361/200912221}, \href
  {http://adsabs.harvard.edu/abs/2009A%26A...503..541L} {503, 541}

\bibitem[\protect\citeauthoryear{{Liu}, {Asplund}, {Yong}, {Mel{\'e}ndez},
  {Ram{\'{\i}}rez}, {Karakas}, {Carlos}  \& {Marino}}{{Liu}
  et~al.}{2016}]{liu/16}
{Liu} F.,  {Asplund} M.,  {Yong} D.,  {Mel{\'e}ndez} J.,  {Ram{\'{\i}}rez} I.,
  {Karakas} A.~I.,  {Carlos} M.,   {Marino} A.~F.,  2016, \mn@doi [\mnras]
  {10.1093/mnras/stw2045}, \href
  {http://adsabs.harvard.edu/abs/2016MNRAS.463..696L} {463, 696}

\bibitem[\protect\citeauthoryear{{Lodders}}{{Lodders}}{2003}]{lodders/03}
{Lodders} K.,  2003, \mn@doi [\apj] {10.1086/375492}, \href
  {http://adsabs.harvard.edu/abs/2003ApJ...591.1220L} {591, 1220}

\bibitem[\protect\citeauthoryear{{Lorenzo-Oliveira} et~al.,}{{Lorenzo-Oliveira}
  et~al.}{2018}]{lorenzo-oliveira/18}
{Lorenzo-Oliveira} D.,  et~al., 2018, \mn@doi [\aap]
  {10.1051/0004-6361/201629294}, \href
  {http://adsabs.harvard.edu/abs/2018A%26A...619A..73L} {619, A73}

\bibitem[\protect\citeauthoryear{{Matsuno}, {Aoki}, {Beers}, {Lee}  \&
  {Honda}}{{Matsuno} et~al.}{2017}]{matsuno/17}
{Matsuno} T.,  {Aoki} W.,  {Beers} T.~C.,  {Lee} Y.~S.,   {Honda} S.,  2017,
  \mn@doi [\aj] {10.3847/1538-3881/aa7a08}, \href
  {http://adsabs.harvard.edu/abs/2017AJ....154...52M} {154, 52}

\bibitem[\protect\citeauthoryear{{Mayor} et~al.,}{{Mayor}
  et~al.}{2003}]{mayor/03}
{Mayor} M.,  et~al., 2003, The Messenger, \href
  {http://adsabs.harvard.edu/abs/2003Msngr.114...20M} {114, 20}

\bibitem[\protect\citeauthoryear{{Mel{\'e}ndez}, {Asplund}, {Gustafsson}  \&
  {Yong}}{{Mel{\'e}ndez} et~al.}{2009}]{melendez/09}
{Mel{\'e}ndez} J.,  {Asplund} M.,  {Gustafsson} B.,   {Yong} D.,  2009, \mn@doi
  [\apjl] {10.1088/0004-637X/704/1/L66}, \href
  {http://adsabs.harvard.edu/abs/2009ApJ...704L..66M} {704, L66}

\bibitem[\protect\citeauthoryear{{Mel{\'e}ndez} et~al.,}{{Mel{\'e}ndez}
  et~al.}{2012}]{melendez/12}
{Mel{\'e}ndez} J.,  et~al., 2012, \mn@doi [\aap] {10.1051/0004-6361/201117222},
  \href {http://adsabs.harvard.edu/abs/2012A%26A...543A..29M} {543, A29}

\bibitem[\protect\citeauthoryear{{Mel{\'e}ndez}, {Schirbel}, {Monroe}, {Yong},
  {Ram{\'{\i}}rez}  \& {Asplund}}{{Mel{\'e}ndez} et~al.}{2014}]{melendeza/14}
{Mel{\'e}ndez} J.,  {Schirbel} L.,  {Monroe} T.~R.,  {Yong} D.,
  {Ram{\'{\i}}rez} I.,   {Asplund} M.,  2014, \mn@doi [\aap]
  {10.1051/0004-6361/201424172}, \href
  {http://adsabs.harvard.edu/abs/2014A%26A...567L...3M} {567, L3}

\bibitem[\protect\citeauthoryear{{Mel{\'e}ndez} et~al.,}{{Mel{\'e}ndez}
  et~al.}{2017}]{melendez/17}
{Mel{\'e}ndez} J.,  et~al., 2017, \mn@doi [\aap] {10.1051/0004-6361/201527775},
  \href {https://ui.adsabs.harvard.edu/#abs/2017A&A...597A..34M} {597, A34}

\bibitem[\protect\citeauthoryear{{Monroe} et~al.,}{{Monroe}
  et~al.}{2013}]{monroe/13}
{Monroe} T.~R.,  et~al., 2013, \mn@doi [\apjl] {10.1088/2041-8205/774/2/L32},
  \href {http://adsabs.harvard.edu/abs/2013ApJ...774L..32M} {774, L32}

\bibitem[\protect\citeauthoryear{{Montalb{\'a}n} \& {Rebolo}}{{Montalb{\'a}n}
  \& {Rebolo}}{2002}]{montalban/02}
{Montalb{\'a}n} J.,  {Rebolo} R.,  2002, \mn@doi [\aap]
  {10.1051/0004-6361:20020338}, \href
  {http://adsabs.harvard.edu/abs/2002A%26A...386.1039M} {386, 1039}

\bibitem[\protect\citeauthoryear{{Naef} et~al.,}{{Naef} et~al.}{2010}]{naef/10}
{Naef} D.,  et~al., 2010, \mn@doi [\aap] {10.1051/0004-6361/200913616}, \href
  {http://adsabs.harvard.edu/abs/2010A%26A...523A..15N} {523, A15}

\bibitem[\protect\citeauthoryear{{Pace} \& {Pasquini}}{{Pace} \&
  {Pasquini}}{2004}]{pace/04}
{Pace} G.,  {Pasquini} L.,  2004, \mn@doi [\aap] {10.1051/0004-6361:20040568},
  \href {http://adsabs.harvard.edu/abs/2004A%26A...426.1021P} {426, 1021}

\bibitem[\protect\citeauthoryear{{Pavlenko}, {Jenkins}, {Ivanyuk}, {Jones},
  {Kaminsky}, {Lyubchik}  \& {Yakovina}}{{Pavlenko} et~al.}{2018}]{pavlenko/18}
{Pavlenko} Y.~V.,  {Jenkins} J.~S.,  {Ivanyuk} O.~M.,  {Jones} H.~R.~A.,
  {Kaminsky} B.~M.,  {Lyubchik} Y.~P.,   {Yakovina} L.~A.,  2018, \mn@doi
  [\aap] {10.1051/0004-6361/201731547}, \href
  {http://adsabs.harvard.edu/abs/2018A%26A...611A..27P} {611, A27}

\bibitem[\protect\citeauthoryear{{Ram{\'{\i}}rez}, {Fish}, {Lambert}  \&
  {Allende Prieto}}{{Ram{\'{\i}}rez} et~al.}{2012}]{ramirez/12}
{Ram{\'{\i}}rez} I.,  {Fish} J.~R.,  {Lambert} D.~L.,   {Allende Prieto} C.,
  2012, \mn@doi [\apj] {10.1088/0004-637X/756/1/46}, \href
  {http://adsabs.harvard.edu/abs/2012ApJ...756...46R} {756, 46}

\bibitem[\protect\citeauthoryear{{Ram{\'{\i}}rez}, {Mel{\'e}ndez}  \&
  {Asplund}}{{Ram{\'{\i}}rez} et~al.}{2014}]{ramirez/14a}
{Ram{\'{\i}}rez} I.,  {Mel{\'e}ndez} J.,   {Asplund} M.,  2014, \mn@doi [\aap]
  {10.1051/0004-6361/201322558}, \href
  {http://adsabs.harvard.edu/abs/2014A%26A...561A...7R} {561, A7}

\bibitem[\protect\citeauthoryear{{Ryan}, {Norris}  \& {Beers}}{{Ryan}
  et~al.}{1999}]{ryan/99}
{Ryan} S.~G.,  {Norris} J.~E.,   {Beers} T.~C.,  1999, \mn@doi [\apj]
  {10.1086/307769}, \href {http://adsabs.harvard.edu/abs/1999ApJ...523..654R}
  {523, 654}

\bibitem[\protect\citeauthoryear{{Sandquist}, {Dokter}, {Lin}  \&
  {Mardling}}{{Sandquist} et~al.}{2002}]{sandquist/02}
{Sandquist} E.~L.,  {Dokter} J.~J.,  {Lin} D.~N.~C.,   {Mardling} R.~A.,  2002,
  \mn@doi [\apj] {10.1086/340452}, \href
  {http://adsabs.harvard.edu/abs/2002ApJ...572.1012S} {572, 1012}

\bibitem[\protect\citeauthoryear{{Schirbel} et~al.,}{{Schirbel}
  et~al.}{2015}]{schirbel/15}
{Schirbel} L.,  et~al., 2015, \mn@doi [\aap] {10.1051/0004-6361/201527303},
  \href {http://adsabs.harvard.edu/abs/2015A%26A...584A.116S} {584, A116}

\bibitem[\protect\citeauthoryear{{Sneden}}{{Sneden}}{1973}]{sneden/73}
{Sneden} C.~A.,  1973, PhD thesis, THE UNIVERSITY OF TEXAS AT AUSTIN.

\bibitem[\protect\citeauthoryear{{Spada}, {Demarque}, {Kim}, {Boyajian}  \&
  {Brewer}}{{Spada} et~al.}{2017}]{spada/16}
{Spada} F.,  {Demarque} P.,  {Kim} Y.-C.,  {Boyajian} T.~S.,   {Brewer} J.~M.,
  2017, \mn@doi [\apj] {10.3847/1538-4357/aa661d}, \href
  {http://adsabs.harvard.edu/abs/2017ApJ...838..161S} {838, 161}

\bibitem[\protect\citeauthoryear{{Spina} et~al.,}{{Spina}
  et~al.}{2018}]{spina/18}
{Spina} L.,  et~al., 2018, \mn@doi [\mnras] {10.1093/mnras/stx2938}, \href
  {http://adsabs.harvard.edu/abs/2018MNRAS.474.2580S} {474, 2580}

\bibitem[\protect\citeauthoryear{{Spite} \& {Spite}}{{Spite} \&
  {Spite}}{1982}]{spite/spite/82}
{Spite} F.,  {Spite} M.,  1982, \aap, \href
  {http://adsabs.harvard.edu/abs/1982A%26A...115..357S} {115, 357}

\bibitem[\protect\citeauthoryear{{Takeda}, {Honda}, {Kawanomoto}, {Ando}  \&
  {Sakurai}}{{Takeda} et~al.}{2010}]{takeda/10}
{Takeda} Y.,  {Honda} S.,  {Kawanomoto} S.,  {Ando} H.,   {Sakurai} T.,  2010,
  \mn@doi [\aap] {10.1051/0004-6361/200913897}, \href
  {http://adsabs.harvard.edu/abs/2010A%26A...515A..93T} {515, A93}

\bibitem[\protect\citeauthoryear{{Th{\'e}ado} \& {Vauclair}}{{Th{\'e}ado} \&
  {Vauclair}}{2012}]{theado/vauclair/12}
{Th{\'e}ado} S.,  {Vauclair} S.,  2012, \mn@doi [\apj]
  {10.1088/0004-637X/744/2/123}, \href
  {http://adsabs.harvard.edu/abs/2012ApJ...744..123T} {744, 123}

\bibitem[\protect\citeauthoryear{{Th{\'e}venin}, {Oreshina}, {Baturin},
  {Gorshkov}, {Morel}  \& {Provost}}{{Th{\'e}venin} et~al.}{2017}]{thevenin/17}
{Th{\'e}venin} F.,  {Oreshina} A.~V.,  {Baturin} V.~A.,  {Gorshkov} A.~B.,
  {Morel} P.,   {Provost} J.,  2017, \mn@doi [\aap]
  {10.1051/0004-6361/201629385}, \href
  {https://ui.adsabs.harvard.edu/#abs/2017A&A...598A..64T} {598, A64}

\bibitem[\protect\citeauthoryear{{Tucci Maia}, {Mel{\'e}ndez}, {Castro},
  {Asplund}, {Ram{\'{\i}}rez}, {Monroe}, {do Nascimento}  \& {Yong}}{{Tucci
  Maia} et~al.}{2015}]{tucci/15}
{Tucci Maia} M.,  {Mel{\'e}ndez} J.,  {Castro} M.,  {Asplund} M.,
  {Ram{\'{\i}}rez} I.,  {Monroe} T.~R.,  {do Nascimento} Jr. J.~D.,   {Yong}
  D.,  2015, \mn@doi [\aap] {10.1051/0004-6361/201425357}, \href
  {http://adsabs.harvard.edu/abs/2015A%26A...576L..10T} {576, L10}

\bibitem[\protect\citeauthoryear{{Xiong} \& {Deng}}{{Xiong} \&
  {Deng}}{2009}]{xiong/deng/09}
{Xiong} D.~R.,  {Deng} L.,  2009, \mn@doi [\mnras]
  {10.1111/j.1365-2966.2009.14581.x}, \href
  {http://adsabs.harvard.edu/abs/2009MNRAS.395.2013X} {395, 2013}

\bibitem[\protect\citeauthoryear{{Yan} et~al.,}{{Yan} et~al.}{2018}]{yan/18}
{Yan} H.-L.,  et~al., 2018, \mn@doi [Nature Astronomy]
  {10.1038/s41550-018-0544-7}, \href
  {http://adsabs.harvard.edu/abs/2018NatAs...2..790Y} {2, 790}

\bibitem[\protect\citeauthoryear{{Yi}, {Demarque}, {Kim}, {Lee}, {Ree},
  {Lejeune}  \& {Barnes}}{{Yi} et~al.}{2001}]{yi/01}
{Yi} S.,  {Demarque} P.,  {Kim} Y.-C.,  {Lee} Y.-W.,  {Ree} C.~H.,  {Lejeune}
  T.,   {Barnes} S.,  2001, \mn@doi [\apjs] {10.1086/321795}, \href
  {http://adsabs.harvard.edu/abs/2001ApJS..136..417Y} {136, 417}

\bibitem[\protect\citeauthoryear{{Zahn}}{{Zahn}}{1994}]{zahn/94}
{Zahn} J.-P.,  1994, \aap, \href
  {http://adsabs.harvard.edu/abs/1994A%26A...288..829Z} {288, 829}

\bibitem[\protect\citeauthoryear{{do Nascimento} Jr. et~al.,}{{do Nascimento}
  et~al.}{2014}]{donascimento/14}
{do Nascimento} Jr. J.-D.,  et~al., 2014, \mn@doi [\apjl]
  {10.1088/2041-8205/790/2/L23}, \href
  {http://adsabs.harvard.edu/abs/2014ApJ...790L..23D} {790, L23}

\bibitem[\protect\citeauthoryear{{dos Santos} et~al.,}{{dos Santos}
  et~al.}{2016}]{dos_santos/16}
{dos Santos} L.~A.,  et~al., 2016, \mn@doi [\aap]
  {10.1051/0004-6361/201628558}, \href
  {http://adsabs.harvard.edu/abs/2016A%26A...592A.156D} {592, A156}

\bibitem[\protect\citeauthoryear{{dos Santos} et~al.,}{{dos Santos}
  et~al.}{2017}]{dos_santos/17}
{dos Santos} L.~A.,  et~al., 2017, \mn@doi [\mnras] {10.1093/mnras/stx2199},
  \href {http://adsabs.harvard.edu/abs/2017MNRAS.472.3425D} {472, 3425}

\makeatother
\end{thebibliography}




\appendix



\begin{table*}
\caption{Li abundances, ages, masses, and stellar parameters.}             
\label{table}      
\begin{center}          
\begin{tabular}{l c c c c c c r  r}     
\hline\hline       
\noalign{\smallskip}
Star & \multicolumn{1}{c}{\begin{tabular}[c]{@{}c@{}}$A$(Li) LTE\\(dex)\end{tabular}} &
\multicolumn{1}{c}{\begin{tabular}[c]{@{}c@{}}$A$(Li) NLTE\\(dex)\end{tabular}} &
\multicolumn{1}{c}{\begin{tabular}[c]{@{}c@{}}Age$^{\ast}$ \\ (Gyr)\end{tabular}} & \multicolumn{1}{c}{\begin{tabular}[c]{@{}c@{}}Mass$^{\ast}$ \\ (M$_{\odot}$) \end{tabular}} & \multicolumn{1}{c}{\begin{tabular}[c]{@{}c@{}} T$^{\ast}_{\mathrm {eff}}$ \\ (K) \end{tabular}} & \multicolumn{1}{c}{\begin{tabular}[c]{@{}c@{}} log \textit{g}$^{\ast}$ \\ (dex) \end{tabular}} & \multicolumn{1}{c}{\begin{tabular}[c]{@{}c@{}}[Fe/H]$^{\ast}$ \\ (dex)\end{tabular}} & Notes\\


\hline 
\noalign{\smallskip}

HIP 1954    &   1.340$^{+0.028}_{-0.061}$    &          1.380$^{+0.028}_{-0.061}$    &     4.80$^{+0.30}_{-0.80}$       & 0.970 &   5720  &  4.46  &  -0.090   &                            \\
HIP 3203    &   2.450$^{+0.005}_{-0.005}$    &          2.452$^{+0.005}_{-0.005}$    &     $\leq$0.50                 & 1.038 &   5868  &  4.54  &  -0.050  &                            \\
HIP 4909    &   2.410$^{+0.005}_{-0.016}$    &          2.424$^{+0.005}_{-0.016}$    &     $\leq$0.60                 & 1.055 &   5861  &  4.50  &  0.048   &                            \\
HIP 5301    &   $\leq$0.910                  &          $\leq$0.952                  &     7.30$^{+0.40}_{-0.50}$       & 0.960 &   5723  &  4.40  &  -0.074  &  Exoplanet detected$^{(a)}$\\
HIP 6407    &   1.770$^{+0.014}_{-0.028}$    &          1.800$^{+0.014}_{-0.028}$    &     1.90$^{+0.70}_{-0.70}$       & 1.004 &   5775  &  4.51  &  -0.058  &  Spectroscopic binary$^{(b)}$        \\
HIP 7585    &   1.790$^{+0.008}_{-0.011}$    &          1.829$^{+0.008}_{-0.011}$    &     3.50$^{+0.30}_{-0.50}$       & 1.043 &   5822  &  4.45  &  0.083   &                            \\
HIP 8507    &   1.530$^{+0.042}_{-0.030}$    &          1.570$^{+0.042}_{-0.030}$    &     4.90$^{+0.40}_{-0.50}$       & 0.961 &   5717  &  4.46  &  -0.099  &                            \\
HIP 9349    &   2.010$^{+0.010}_{-0.011}$    &          2.036$^{+0.010}_{-0.011}$    &     0.60$^{+0.40}_{-0.30}$       & 1.036 &   5818  &  4.52  &  -0.006  &                            \\
HIP 10175   &   1.690$^{+0.014}_{-0.022}$    &          1.730$^{+0.014}_{-0.022}$    &     3.10$^{+0.40}_{-0.30}$       & 0.990 &   5719  &  4.49  &  -0.028  &                            \\
HIP 10303   &   1.490$^{+0.014}_{-0.014}$    &          1.540$^{+0.014}_{-0.014}$    &     5.90$^{+0.40}_{-0.40}$       & 1.011 &   5712  &  4.40  &  0.104   &                            \\
HIP 11915   &   1.570$^{+0.010}_{-0.014}$    &          1.604$^{+0.010}_{-0.014}$    &     3.60$^{+0.50}_{-0.70}$       & 0.993 &   5769  &  4.48  &  -0.067  &  Exoplanet detected$^{(c)}$   \\
HIP 14501   &   $\leq$0.220                  &          $\leq$0.260                  &     8.80$^{+0.30}_{-0.30}$       & 0.979 &   5738  &  4.31  &  -0.153  &  Spectroscopic binary$^{(b)}$        \\
HIP 14614   &   1.600$^{+0.014}_{-0.014}$    &          1.628$^{+0.014}_{-0.014}$    &     4.70$^{+0.40}_{-0.60}$       & 0.986 &   5803  &  4.45  &  -0.109  &                            \\
HIP 15527   &   0.640$^{+0.141}_{-0.224}$    &          0.676$^{+0.141}_{-0.224}$    &     7.70$^{+0.40}_{-0.30}$       & 0.986 &   5779  &  4.34  &  -0.064  &  Exoplanet detected$^{(d)}$\\
HIP 18844   &   0.670$^{+0.180}_{-0.269}$    &          0.720$^{+0.180}_{-0.269}$    &     7.00$^{+0.30}_{-0.40}$       & 0.997 &   5734  &  4.37  &  0.014   &  Spectroscopic binary$^{(b)}$       \\
HIP 22263   &   2.370$^{+0.005}_{-0.010}$    &          2.383$^{+0.005}_{-0.010}$    &     0.80$^{+0.30}_{-0.40}$       & 1.052 &   5870  &  4.54  &  0.037   &                            \\
HIP 25670   &   1.110$^{+0.050}_{-0.050}$    &          1.160$^{+0.050}_{-0.050}$    &     5.10$^{+0.30}_{-0.30}$       & 1.010 &   5760  &  4.42  &  0.054   &                            \\
HIP 28066   &   0.710$^{+0.054}_{-0.058}$    &          0.745$^{+0.054}_{-0.058}$    &     8.80$^{+0.30}_{-0.30}$       & 0.989 &   5742  &  4.30  &  -0.147  &                            \\
HIP 29432   &   1.210$^{+0.036}_{-0.022}$    &          1.245$^{+0.036}_{-0.022}$    &     5.20$^{+0.40}_{-0.40}$       & 0.969 &   5762  &  4.45  &  -0.112  &                            \\
HIP 30037   &   0.740$^{+0.141}_{-0.224}$    &          0.790$^{+0.141}_{-0.224}$    &     6.70$^{+0.50}_{-0.50}$       & 0.960 &   5666  &  4.42  &  0.007  &  Spectroscopic binary$^{(b)}$       \\
HIP 30158   &   0.670$^{+0.100}_{-0.197}$    &          0.720$^{+0.100}_{-0.197}$    &     7.90$^{+0.30}_{-0.30}$       & 0.963 &   5678  &  4.37  &  -0.004  &                            \\
HIP 30476   &   $\leq$0.270                  &          $\leq$0.315                  &     9.00$^{+0.30}_{-0.30}$       & 0.990 &   5709  &  4.28  &  -0.033  &                  \\
HIP 30502   &   0.950$^{+0.099}_{-0.094}$    &          0.990$^{+0.099}_{-0.094}$    &     7.00$^{+0.40}_{-0.10}$       & 0.965 &   5731  &  4.40  &  -0.057  &                            \\
HIP 33094   &   0.620$^{+0.166}_{-0.089}$    &          0.678$^{+0.166}_{-0.089}$    &     8.90$^{+0.30}_{-0.30}$       & 1.064 &   5629  &  4.11  &  0.023   &                            \\
HIP 34511   &   1.730$^{+0.005}_{-0.022}$    &          1.756$^{+0.005}_{-0.022}$    &     4.00$^{+0.50}_{-0.40}$       & 0.998 &   5812  &  4.45  &  -0.091  &                            \\
HIP 36512   &   1.200$^{+0.036}_{-0.036}$    &          1.236$^{+0.036}_{-0.036}$    &     5.90$^{+0.40}_{-0.50}$       & 0.957 &   5744  &  4.45  &  -0.126  &                            \\
HIP 36515   &   2.680$^{+0.005}_{-0.005}$    &          2.667$^{+0.005}_{-0.005}$    &     0.50$^{+0.30}_{-0.30}$       & 1.031 &   5855  &  4.56  &  -0.029  &                            \\
HIP 38072   &   1.620$^{+0.054}_{-0.028}$    &          1.656$^{+0.054}_{-0.028}$    &     1.00$^{+0.80}_{-0.50}$       & 1.063 &   5860  &  4.51  &  0.085   &                            \\
HIP 40133   &   1.480$^{+0.011}_{-0.021}$    &          1.530$^{+0.011}_{-0.021}$    &     5.40$^{+0.30}_{-0.30}$       & 1.040 &   5745  &  4.37  &  0.116   &                            \\
HIP 41317   &   0.690$^{+0.081}_{-0.186}$    &          0.733$^{+0.081}_{-0.186}$    &     7.70$^{+0.30}_{-0.30}$       & 0.960 &   5706  &  4.39  &  -0.081  &                            \\
HIP 42333   &   2.250$^{+0.006}_{-0.006}$    &          2.280$^{+0.006}_{-0.006}$    &     1.00$^{+0.70}_{-0.40}$       & 1.069 &   5846  &  4.50  &  0.132   &                            \\
HIP 43297   &   1.590$^{+0.014}_{-0.010}$    &          1.640$^{+0.014}_{-0.010}$    &     1.80$^{+0.50}_{-0.40}$       & 1.014 &   5705  &  4.51  &  0.082   &                           \\
HIP 44713   &   0.590$^{+0.141}_{-0.355}$    &          0.638$^{+0.141}_{-0.355}$    &     7.70$^{+0.30}_{-0.30}$       & 1.029 &   5759  &  4.28  &  0.063  &                            \\
HIP 44935   &   0.980$^{+0.057}_{-0.090}$    &          1.020$^{+0.057}_{-0.090}$    &     6.60$^{+0.30}_{-0.40}$       & 1.009 &   5771  &  4.37  &  0.038   &                            \\
HIP 44997   &   1.140$^{+0.043}_{-0.036}$    &          1.184$^{+0.043}_{-0.036}$    &     6.60$^{+0.40}_{-0.40}$       & 0.970 &   5728  &  4.41  &  -0.012  &                            \\
HIP 49756   &   1.410$^{+0.014}_{-0.025}$    &          1.450$^{+0.014}_{-0.025}$    &     4.50$^{+0.30}_{-0.40}$       & 1.010 &   5789  &  4.44  &  0.023   &                            \\
HIP 54102   &   2.170$^{+0.011}_{-0.010}$    &          2.191$^{+0.011}_{-0.010}$    &     0.70$^{+0.40}_{-0.40}$       & 1.047 &   5845  &  4.51  &  0.011   &  Spectroscopic binary$^{(b)}$        \\
HIP 54287   &   1.860$^{+0.007}_{-0.011}$    &          1.911$^{+0.007}_{-0.011}$    &     6.50$^{+0.30}_{-0.40}$       & 1.024 &   5714  &  4.34  &  0.107   &                            \\
HIP 54582   &   1.620$^{+0.011}_{-0.022}$    &          1.640$^{+0.011}_{-0.022}$    &     6.90$^{+0.30}_{-0.30}$       & 1.034 &   5883  &  4.28  &  -0.096  &  Spectroscopic binary$^{(b)}$       \\
HIP 62039   &   0.760$^{+0.067}_{-0.184}$    &          0.814$^{+0.067}_{-0.184}$    &     6.20$^{+0.40}_{-0.30}$       & 1.040 &   5742  &  4.34  &  0.104   &  Spectroscopic binary$^{(b)}$       \\
HIP 64150   &   $\leq$0.440                  &          $\leq$0.490                  &     6.40$^{+0.30}_{-0.30}$       & 1.010 &   5747  &  4.37  &  0.049   &  Spectroscopic binary$^{(b)}$     \\
HIP 64673   &   1.780$^{+0.010}_{-0.036}$    &          1.799$^{+0.010}_{-0.036}$    &     6.00$^{+0.40}_{-0.40}$       & 1.068 &   5912  &  4.29  &  -0.017  &                 \\
HIP 64713   &   1.420$^{+0.014}_{-0.036}$    &          1.454$^{+0.014}_{-0.036}$    &     5.30$^{+0.50}_{-0.60}$       & 0.989 &   5788  &  4.44  &  -0.043  &                            \\
HIP 65708   &   0.710$^{+0.144}_{-0.089}$    &          0.750$^{+0.144}_{-0.089}$    &     9.00$^{+0.30}_{-0.30}$       & 1.009 &   5746  &  4.22  &  -0.063  &  Spectroscopic binary$^{(b)}$       \\
HIP 68468   &   1.460$^{+0.014}_{-0.071}$    &          1.497$^{+0.014}_{-0.071}$    &     5.50$^{+0.30}_{-0.40}$       & 1.064 &   5845  &  4.33  &  0.071   &  Exoplanet detected$^{(e)}$\\
HIP 69645   &   1.040$^{+0.057}_{-0.143}$    &          1.080$^{+0.057}_{-0.143}$    &     5.70$^{+0.30}_{-0.90}$       & 0.986 &   5751  &  4.44  &  -0.026  &                            \\
HIP 72043   &   1.030$^{+0.100}_{-0.076}$    &          1.060$^{+0.100}_{-0.076}$    &     6.20$^{+0.40}_{-0.30}$       & 1.026 &   5845  &  4.34  &  -0.026  &  Spectroscopic binary$^{(b)}$      \\
HIP 73241   &   $\leq$0.180                  &          $\leq$0.240                  &     8.90$^{+0.30}_{-0.30}$       & 1.031 &   5661  &  4.22  &  0.092   &  Spectroscopic binary$^{(b)}$     \\
HIP 73815   &   0.870$^{+0.099}_{-0.122}$    &          0.910$^{+0.099}_{-0.122}$    &     7.20$^{+0.30}_{-0.30}$       & 1.011 &   5790  &  4.33  &  0.023   &                            \\
HIP 74389   &   2.060$^{+0.005}_{-0.013}$    &          2.090$^{+0.005}_{-0.013}$    &     3.90$^{+0.30}_{-0.60}$       & 1.049 &   5845  &  4.44  &  0.083   &                            \\
HIP 74432   &   0.590$^{+0.156}_{-0.112}$    &          0.640$^{+0.156}_{-0.112}$    &     8.60$^{+0.30}_{-0.30}$       & 1.056 &   5679  &  4.17  &  0.048   &                \\
HIP 76114   &   0.910$^{+0.064}_{-0.085}$    &          0.950$^{+0.064}_{-0.085}$    &     6.60$^{+0.30}_{-0.30}$       & 0.980 &   5740  &  4.41  &  -0.024  &                            \\
HIP 77052   &   1.510$^{+0.018}_{-0.022}$    &          1.564$^{+0.018}_{-0.022}$    &     4.50$^{+1.10}_{-0.40}$       & 0.985 &   5687  &  4.45  &  0.051   &  Visual binary$^{(b)}$              \\
HIP 77883   &   0.660$^{+0.061}_{-0.114}$    &          0.710$^{+0.061}_{-0.114}$    &     7.60$^{+0.30}_{-0.40}$       & 0.970 &   5699  &  4.34  &  0.017   &                            \\
HIP 79578   &   1.940$^{+0.005}_{-0.006}$    &          1.970$^{+0.005}_{-0.006}$    &     2.40$^{+0.60}_{-0.40}$       & 1.031 &   5810  &  4.47  &  0.048   &  Spectroscopic binary$^{(b)}$      \\
HIP 79672   &   1.570$^{+0.011}_{-0.011}$    &          1.608$^{+0.011}_{-0.011}$    &     4.20$^{+0.30}_{-0.50}$       & 1.022 &   5808  &  4.44  &  0.041   &                            \\
HIP 79715   &   1.050$^{+0.140}_{-0.085}$    &          1.080$^{+0.140}_{-0.085}$    &     6.20$^{+0.30}_{-0.40}$       & 1.000 &   5816  &  4.38  &  -0.037  &                            \\

\noalign{\smallskip}
\hline 
\end{tabular}
\end{center}
Table 1 continued
\end{table*}

\begin{table*}
\contcaption{}             
\label{table_cont}      
\begin{center}          
\begin{tabular}{l r r c c c c r  r}     
\hline\hline       
\noalign{\smallskip}
Star & \multicolumn{1}{c}{\begin{tabular}[c]{@{}c@{}}$A$(Li) LTE\\(dex)\end{tabular}} &
\multicolumn{1}{c}{\begin{tabular}[c]{@{}c@{}}$A$(Li) NLTE\\(dex)\end{tabular}} &
\multicolumn{1}{c}{\begin{tabular}[c]{@{}c@{}}Age$^{\ast}$ \\ (Gyr)\end{tabular}} & \multicolumn{1}{c}{\begin{tabular}[c]{@{}c@{}}Mass$^{\ast}$ \\ (M$_{\odot}$) \end{tabular}} & \multicolumn{1}{c}{\begin{tabular}[c]{@{}c@{}} T$^{\ast}_{\mathrm {eff}}$ \\ (K) \end{tabular}} & \multicolumn{1}{c}{\begin{tabular}[c]{@{}c@{}} log \textit{g}$^{\ast}$ \\ (dex) \end{tabular}} & \multicolumn{1}{c}{\begin{tabular}[c]{@{}c@{}}[Fe/H]$^{\ast}$ \\ (dex)\end{tabular}} & Notes\\


\hline 
\noalign{\smallskip}

HIP 81746   &   0.590$^{+0.242}_{-0.112}$    &          0.630$^{+0.242}_{-0.112}$    &     8.10$^{+0.30}_{-0.30}$       & 0.960 &   5715  &  4.37  &  -0.091  &  Spectroscopic binary$^{(b)}$      \\
HIP 83276   &   1.670$^{+0.018}_{-0.009}$    &          1.690$^{+0.018}_{-0.009}$    &     7.40$^{+0.30}_{-0.30}$       & 1.033 &   5886  &  4.24  &  -0.093  &  Spectroscopic binary$^{(b)}$              \\
HIP 85042   &   0.480$^{+0.124}_{-0.040}$    &          0.533$^{+0.124}_{-0.040}$    &     7.80$^{+0.30}_{-0.30}$       & 0.970 &   5685  &  4.35  &  0.030   &                            \\
HIP 87769   &   1.570$^{+0.064}_{-0.030}$    &          1.610$^{+0.064}_{-0.030}$    &     5.00$^{+0.40}_{-1.00}$       & 1.039 &   5828  &  4.40  &  0.072   &  Spectroscopic binary$^{(b)}$      \\
HIP 89650   &   1.380$^{+0.024}_{-0.081}$    &          1.409$^{+0.024}_{-0.081}$    &     4.30$^{+0.70}_{-0.30}$       & 1.027 &   5851  &  4.42  &  -0.015  &                            \\
HIP 95962   &   1.230$^{+0.021}_{-0.106}$    &          1.269$^{+0.021}_{-0.106}$    &     6.00$^{+0.40}_{-0.30}$       & 1.010 &   5805  &  4.38  &  0.029   &                \\
HIP 96160   &   1.720$^{+0.007}_{-0.029}$    &          1.750$^{+0.007}_{-0.029}$    &     2.60$^{+0.40}_{-0.50}$       & 1.012 &   5798  &  4.48  &  -0.036  &                            \\
HIP 101905  &   2.120$^{+0.010}_{-0.014}$    &          2.145$^{+0.010}_{-0.014}$    &     1.20$^{+0.30}_{-0.30}$       & 1.080 &   5906  &  4.50  &  0.088  &                            \\
HIP 102040  &   2.160$^{+0.010}_{-0.010}$    &          2.170$^{+0.010}_{-0.010}$    &     2.40$^{+0.40}_{-0.40}$       & 1.020 &   5853  &  4.48  &  -0.079  &                            \\
HIP 102152  &   0.580$^{+0.212}_{-0.224}$    &          0.630$^{+0.212}_{-0.224}$    &     8.60$^{+0.30}_{-0.40}$       & 0.978 &   5718  &  4.33  &  -0.016  &                            \\
HIP 104045  &   1.510$^{+0.064}_{-0.030}$    &          1.550$^{+0.064}_{-0.030}$    &     4.10$^{+0.90}_{-0.30}$       & 1.027 &   5826  &  4.41  &  0.051   &                            \\
HIP 105184  &   2.230$^{+0.011}_{-0.014}$    &          2.247$^{+0.011}_{-0.014}$    &     0.60$^{+0.50}_{-0.30}$       & 1.050 &   5843  &  4.51  &  0.003   &                            \\
HIP 108158  &   0.560$^{+0.194}_{-0.252}$    &          0.616$^{+0.194}_{-0.252}$    &     8.10$^{+0.30}_{-0.30}$       & 1.021 &   5675  &  4.29  &  0.055  &                            \\
HIP 108468  &   1.100$^{+0.058}_{-0.122}$    &          1.127$^{+0.058}_{-0.122}$    &     7.00$^{+0.30}_{-0.30}$       & 1.006 &   5841  &  4.35  &  -0.096  &                            \\
HIP 109821  &   0.670$^{+0.136}_{-0.234}$    &          0.707$^{+0.136}_{-0.234}$    &     8.90$^{+0.30}_{-0.30}$       & 0.980 &   5747  &  4.31  &  -0.108  &                            \\
HIP 114615  &   1.860$^{+0.014}_{-0.011}$    &          1.886$^{+0.014}_{-0.011}$    &     0.50$^{+1.20}_{-0.30}$       & 1.027 &   5819  &  4.51  &  -0.063  &                            \\
HIP 115577  &   $\leq$0.160                 &          $\leq$0.210                 &     8.80$^{+0.30}_{-0.30}$       & 1.019 &   5694  &  4.26  &  0.013   &                  \\
HIP 116906  &   0.740$^{+0.191}_{-0.112}$    &          0.778$^{+0.191}_{-0.112}$    &     6.70$^{+0.30}_{-0.30}$       & 1.010 &   5790  &  4.37  &  -0.005  &  Exoplanet detected$^{(f)}$ \\
HIP 117367  &   1.420$^{+0.028}_{-0.058}$    &          1.450$^{+0.028}_{-0.058}$    &     5.70$^{+0.30}_{-0.30}$       & 1.040 &   5867  &  4.35  &  0.024   &                            \\
HIP 118115  &   0.920$^{+0.042}_{-0.081}$    &          0.960$^{+0.042}_{-0.081}$    &     8.00$^{+0.30}_{-0.30}$       & 1.013 &   5798  &  4.28  &  -0.036  &                            \\
Sun$^{a}$   &   1.030$^{+0.030}_{-0.020}$    &          1.070$^{+0.030}_{-0.020}$    &     4.6                        & 1.000 &   5777  &  4.44  &  0.000   &                            \\

\noalign{\smallskip}
\hline 
\end{tabular}
\end{center}
\begin{tabular}{l}
\textbf{Notes.} $^{(\ast)}$Data from \cite{spina/18}. $^{(a)}$\cite{naef/10}. $^{(b)}$\cite{dos_santos/17}.  $^{(c)}$\cite{bedell/15}. $^{(d)}$\cite{jones/06}. \\
\hspace{0.95cm} $^{(e)}$\cite{melendez/17}. $^{(f)}$\cite{butler/06}.\\
\end{tabular}

\end{table*}


\bsp	
\label{lastpage}
\end{document}